\documentclass[twocolumn]{emulateapj}  
\usepackage{times}

\usepackage{graphicx,bm,amssymb}
\usepackage{epsfig}
\usepackage{amssymb}
\usepackage{natbib}
\usepackage{pstricks}
\usepackage{psfrag}
\usepackage[colorlinks=true,linkcolor=blue,citecolor=blue]{hyperref}
\usepackage{float}
\usepackage[caption = false]{subfig}
\def\araa{ARA\&A}
\def\apj{ApJ}
\def\apjl{ApJ}
\def\apjs{ApJS}
\def\aap{A\&A}
\def\aaps{A\&AS}
\def\mnras{MNRAS}
\def\nat{Nature}
\def\physrep{Phys.~Rep.}

\newcommand{\be}{\begin{equation}}
\newcommand{\ee}{\end{equation}}
\newcommand{\bary}{\begin{eqnarray}}
\newcommand{\eary}{\end{eqnarray}}

\shorttitle{GRB 170817A}
\shortauthors{Fraija N.}

\begin{document}
\title{Light curves of a shock-breakout material and a relativistic\\ off-axis jet from a Binary Neutron Star system}
\author{N. Fraija$^{1\dagger}$,  A.C. Caligula do E. S. Pedreira$^{1,2}$ and P. Veres$^3$}
\affil{$^1$ Instituto de Astronom\' ia, Universidad Nacional Aut\'onoma de M\'exico, Circuito Exterior, C.U., A. Postal 70-264, 04510 Cd. de M\'exico,  M\'exico.\\
$^2$ Instituto de Matem\'atica, Estat\'isca e F\'isica, Universidade Federal do Rio Grande, Rio Grande 96203-900, Brasil\\
$^3$ Center for Space Plasma and Aeronomic Research (CSPAR), University of Alabama in Huntsville, Huntsville, AL 35899, USA\\
}
\email{$\dagger$nifraija@astro.unam.mx}
\date{\today} 
%
%
\begin{abstract}
Binary neutron star mergers are believed to eject significant masses with a diverse range of velocities.  Once these ejected materials begin to be decelerated by a homogeneous medium,  relativistic electrons are mainly cooled down by synchrotron radiation, generating a multiwavelength long-lived afterglow.   Analytic and numerical methods illustrate that the outermost matter, the merger shock-breakout material, can be parametrized by power-law velocity distributions  $\propto \left(\beta_{\rm c}\Gamma \right)^{-\alpha_s}$.  Considering that the shock-breakout  material is moving on-axis towards the observer and the relativistic jet off-axis, we compute the light curves during the relativistic and the lateral expansion phase.   As a particular case, we successfully describe the X-ray, optical and radio light curves alongside the spectral energy distribution from the recently discovered gravitational-wave transient GW170817,  when the merger shock-breakout material moves with mildly relativistic velocities near-Newtonian phase and the jet with relativistic velocities.   Future electromagnetic counterpart observations of this binary system could be able to evaluate different properties of these light curves.\\  
\end{abstract}
\section{Introduction}
Binary neutron star (NS) mergers are thought to be natural candidates for gravitational waves (GWs), short gamma-ray bursts (sGRBs), mass ejections producing   delayed radio emissions and  isotropic quasi-thermal optical/infrared  counterparts, so-called kilonova/macronova \citep[for reviews, see][]{2007PhR...442..166N, 2014ARA&A..52...43B, 2017LRR....20....3M}.   Kilonova/macronova is related to a neutron-rich mass ejection ($\sim 10^{-4} - 10^{-2}\,M_\odot$) which presents a rapid neutron capture process (r-process) nucleosynthesis  \citep{1974ApJ...192L.145L, 1976ApJ...210..549L}. This process  synthesizes heavy and unstable nuclei - such as gold and platinum - and consequently heats rapidly the merger ejecta by the radioactive decay energy \citep{1998ApJ...507L..59L, 2005ApJ...634.1202R, 2010MNRAS.406.2650M, 2013ApJ...774...25K, 2017LRR....20....3M}.  SGRBs, with duration less than 2 s, originate from internal collisions or magnetic dissipation within the beamed and relativistic outflow.  Delayed radio emission is expected from the interactions of the ejected materials with the circumburst medium \citep{2011Natur.478...82N, 2013MNRAS.430.2121P, 2015MNRAS.450.1430H}.  \cite{2014MNRAS.437L...6K} proposed the possibility of detecting X-ray, optical and radio fluxes from an ejected ultrarelativistic material decelerated early  (from seconds to days) by the interstellar medium (ISM). \\   
\\
The gravitational-wave transient GW170817, associated with a binary NS system with a merger time of 12:41:04 UTC, 2017 August 17,  was detected by LIGO and Virgo experiments \citep{PhysRevLett.119.161101,2041-8205-848-2-L12}.  Immediately, GRB 170817A triggered the Gamma-ray Burst Monitor (GBM) onboard Fermi Gamma-ray Space Telescope at 12:41:06 UTC \citep{2017ApJ...848L..14G}. The INTErnational Gamma-Ray Astrophysics Laboratory (INTEGRAL) detected an attenuated $\gamma$-ray flux with $\sim 3\sigma$ \citep{2017ApJ...848L..15S}.  GRB 170817A was followed up  by multiple ground-based telescopes in different bands.   A bright optical i-band flux with magnitude $17.057\pm 0.0018$ mag was detected by the 1-meter Swope telescope at Las Campanas Observatory in Chile after 10.87 hours, followed  by multiple  optical and infrared ground-based telescopes.    The X-ray and radio experiments followed-up  this burst during the subsequent days after the merger without detecting any signal. Finally, GRB 170817A began to be detected in X-rays on the ninth day by Chandra \citep{troja2017a, 2017ATel11037....1M, 2018ATel11242....1H}, in the radio (3 and 6 GHz) bands on day  nineteenth by Very Large Array and Atacama Large Millimeter/submillimeter Array  \citep[VLA and ALMA, respectively;][]{2017ApJ...848L..21A} and in the optical band after $\sim$ 110 days \citep[see;][]{2018arXiv180103531M}.  It is worth highlighting that although optical afterglow was detected from the month of December, quasi-thermal optical emission associated to the kilonova was detected early; around 11 hours after the GBM trigger.  The host galaxy associated with this event, NGC 4993, was located at a distance of ($z\simeq 0.01$) 40 Mpc \citep{2017arXiv171005452C,2017ApJ...848L..20M}.\\
Whereas several authors have associated the early $\gamma$-ray photons to different emission mechanisms \citep{2017arXiv171005896G, 2017arXiv171005897B, 2017arXiv171100243K}, the  X-ray, optical and radio afterglow have been related to the synchrotron forward-shock radiation, when the relativistic off-axis jet and/or cocoon are decelerated in an homogeneous low density medium in the range  $10^{-4}$ - $10^{-2}\,{\rm cm^{-3}}$ \citep{2017arXiv171008514F, 2017ApJ...848L..34M, 2017arXiv171005905I, 2017arXiv171111573M, 2017arXiv171203237L, 2017arXiv171006421G, 2017ApJ...848L..21A, 2017ApJ...848L..20M, 2017Sci...358.1559K, 2017arXiv171005822P, 2018ApJ...853L..13W,2018arXiv180106516T}.\\

In this paper,  we derive the forward shock dynamics and the synchrotron light curves from the outermost shock-breakout material and the relativistic off-axis jet from a binary NS system.  As a particular case,  we explain the electromagnetic counterpart detected in the GW170817 transient. Our proposed model uses general arguments of the synchrotron afterglow theory introduced in \cite{1998ApJ...497L..17S} and  the results obtained numerically in \cite{2001ApJ...551..946T} about the ejected masses from GRB progenitors.  This paper is arranged as follows: In Section 2 and 3 we  derive the synchrotron forward-shock model generated by the deceleration of  the shock-breakout material and the relativistic off-axis jet  which were launched from a binary NS merger.  In Section 4,  we present the data used in this work and describe the  electromagnetic counterpart  of GW170817 with our synchrotron forward-shock model.  In section 5, we discuss the results and  present our conclusions.\\
The convention $Q_{\rm x}=Q/10^{\rm x}$ in c.g.s. units and $\hbar$=c=1 in natural units will be used.  The values of cosmological parameters $H_0=$ 71 km s$^{-1}$ Mpc$^{-1}$, $\Omega_m=0.27$, $\Omega_\lambda=0.73$ are adopted \citep{2003ApJS..148..175S}.  
\section{The merger shock-breakout material} 
\subsection{Properties and Considerations}
The breakout burst signal properties depend on the mass ($M_{\rm m}$), radius ($R_{\rm m}$) and velocity ($\beta_{\rm b}$) of the merger remnant.  Right after the binary NS merger takes place, a shock formed at the interface between the NSs  is initially  launched from the core towards the crust at sub-relativistic velocities \citep[$\beta_{\rm b,in}\sim0.25$;][]{2014MNRAS.437L...6K, 2015MNRAS.446.1115M}. At this early phase, the shocked material cannot escape from the merged remnant because the initial velocity is less than the escape velocity  {\small $\beta_{\rm b,es}\simeq 0.83\,   \left(\frac{M_{\rm m}}{3\,M_\odot}\right)^{\frac12} \left(\frac{15\, {\rm km}}{R_{\rm m}}\right)^{\frac12} $}.\\
\\
Although  the shock velocity increases as the crust density decreases ($\rho\propto r^{-s}$ with $s\simeq0.187$ for a polytropic index $n_p=3$), it will eventually reach the escape velocity. Once the shock has reached half of the escape velocity, it can leave the binary NS merger by converting thermal energy into kinetic energy.  At this moment,  the shock velocity has increased by a factor of $0.5\beta_{\rm b,esc}/\beta_{\rm b,ini}\simeq \frac{3}{2}$ and the crust density has decreased by $\simeq \left(\frac32\right)^{-1/s}$. The ejected mass can be estimated as
{\small
\bary
M_{b}&\simeq& 5\times 10^{-5 }\, M_\odot \left(\frac{M_c}{10^{-2} M_\odot}\right)\left(\frac{R_{\rm b}}{1\, {\rm km}} \right)   \left(\frac{15\, {\rm km}}{R_{\rm b}} \right)\,\left(\frac{\beta_{\rm b,ini}}{0.25}\right)^{\frac{5}{7}}\cr
&& \hspace{5.2cm}\times \left(\frac{\beta_{\rm b,esc}}{0.83}\right)^{-\frac{5}{7}}\,,
\eary
}
where  $M_c$ is the crust mass. Once the merger material has moved for long enough to achieve the near-Newtonian phase,  the decelerated material propagates adiabatically with the same effective polytropic index \citep{2010A&A...520L...3M, 2010MNRAS.403..300V}. \\
\cite{2001ApJ...551..946T} investigated the acceleration of the shock waves to relativistic and sub-relativistic velocities in the outer matter of an explosion.  They found that when the energy of the explosion was concentrated in the outermost ejecta,  the blast wave could generate a strong electromagnetic emission. Furthermore, the equivalent kinetic energy of the outermost matter can be expressed as a power-law velocity distribution given by 
{\small
\begin{eqnarray}
\label{beta}
E_{\rm k, c}(\gtrsim\beta_{\rm c}\Gamma_{\rm c}) = \tilde{E} \cases{ 
 \left( \beta_{\rm c}\Gamma_{\rm c} \right)^{-(1.58\gamma_p-1)}   \,\hspace{0.2cm} {\rm for}\,\hspace{0.2cm}(\beta_{\rm c}\Gamma_{\rm c})\gg 1  \,, \cr
 \left( \beta_{\rm c}\Gamma_{\rm c} \right)^{-(5.35\gamma_p-2)}   \,\hspace{0.2cm} {\rm for}\,\hspace{0.2cm}(\beta_{\rm c}\Gamma_{\rm c})\ll 1  \,, \cr
}
\end{eqnarray}
}
where $\tilde{E}$ is the  fiducial energy  and $\gamma_p=1+1/n_p$ with $n_p$  the polytropic index.  For $n_p=3$,  the kinetic equivalent energy  is given by 
\be\label{Ek}
E_{\rm k, c}(\gtrsim\beta_{\rm c} \Gamma_{\rm c}) \simeq \tilde{E}\,\left(\beta_{\rm c}\Gamma_{\rm c}\right)^{-\alpha_s}\,,   
\ee
with $\alpha_s=1.1$ for $\beta_{\rm c}\Gamma_{\rm c}\gg 1$ and $\alpha_s=5.2$ for $\beta_{\rm c}\Gamma_{\rm c}\ll 1$.\\
Once the relativistic shock-breakout material sweeps up enough ISM, the electron population is cooled down emitting synchrotron radiation in the observer's line-of-sight.  Assuming that all the energy is confined within an opening angle $\theta_c\sim 1/\Gamma_{\rm c}$ \citep{2001ApJ...551..946T},  the equivalent kinetic energy associated to the observed region becomes  $\simeq \frac{E_{\rm k,c}(>\Gamma_{\rm c})}{2\Gamma_{\rm c}^2}$  \citep{2018arXiv180109712N}, and therefore from eq. (\ref{Ek}) the observed kinetic energy, $E_{\rm k,c}$, can be written as 
\be\label{Ek_obs}
E_{\rm k, c}(\gtrsim\beta_{\rm c}\Gamma_{\rm c}) \simeq \tilde{E}\,\left(\beta_{\rm c}\Gamma_{\rm c}\right)^{-\delta}\,,   
\ee
with $\delta=\alpha_s+2$.  The factor of $2$ has been absorbed in the fiducial energy $\tilde{E}$. On the other hand,  the electromagnetic energy released in the observer's direction can be estimated through the efficiency $\eta$ and the equivalent kinetic energy $E_{\rm obs,\gamma}\approx \eta\,E_{\rm k, c}(\gtrsim\beta_{\rm c}\Gamma_{\rm c})= \eta\tilde{E}\, \left(\beta_{\rm c}\Gamma_{\rm c}\right)^{-\delta}$.  Therefore, the efficiency of kinetic  to $\gamma$-ray energy conversion can be estimated as 
\be\label{eta}
\eta\simeq  \frac {E_{\rm iso, \gamma}\,\left(\beta_{\rm c} \Gamma_{\rm c} \right)^{{\delta}} } {\,\tilde{E}}\,, 
\ee
where $\tilde{E}$ is a function of total kinetic energy.
\subsection{Analytical model: Synchrotron forward-shock emission}
\subsubsection{Relativistic Phase ($\Gamma_{\rm c} > 1/\theta_{\rm c}$ and $\beta_{\rm c}= 1$)}
The forward shock dynamics when the outflow with a constant equivalent kinetic energy is decelerated by the ISM  has been exhaustively explored \cite[see, e.g.][]{1998ApJ...497L..17S}.   Here, we derive the dynamics of the forward shock when kinetic energy can be described as  a power-law velocity distribution parametrized in accordance with eq. (\ref{Ek}).   Taking into consideration the adiabatic evolution of the shock, the fiducial energy is given by $\tilde{E}=4/3\pi\Gamma_{\rm c}^{2+\delta}R^3n\,m_p$ \citep{1976PhFl...19.1130B, 1997ApJ...489L..37S} where  $n$ is the homogeneous ISM density,  $m_p$ is the proton mass and  $R=\frac{\Gamma_{\rm c}^2\,t}{1+z}$ is the deceleration radius.  The deceleration timescale and bulk Lorentz factor are given by
{\small
\be\label{t_dec}
t_{\rm dec}\simeq 4.2\,{\rm d}\,\left(\frac{1+z}{1.02}\right)\,n^{-\frac13}_{-4}\,\tilde{E}^{\frac13}_{49}\,\Gamma_{\rm c, 0.5}^{-\frac{\delta+8}{3}}\,,
\ee
}
and 
{\small
\be
\Gamma_{\rm c}=3.1\,\left(\frac{1+z}{1.02}\right)^{\frac{3}{\delta+8}}  \,n^{-\frac{1}{\delta+8}}_{-4}\,\tilde{E}^{\frac{1}{\delta+8}}_{49}\,t^{-\frac{3}{\delta+8}}_{\rm  15 d}\,,
\ee
}
respectively.  Hereafter through this section the value of $\alpha_s=3.0$ \citep{2015MNRAS.450.1430H}  is considered.\\
In the synchrotron forward-shock framework, the accelerated electron population is described by  $\gamma_e\geq\gamma_m: N(\gamma_e)\,d\gamma_e \propto \gamma_e^{-p}\,d\gamma_e$, where $p=2.2$ is the electron power index and $\gamma_m$ is the minimum Lorentz factor given by
{\small
\begin{eqnarray}\label{gammam}
\gamma_{\rm m}&=& 68.4\, \epsilon_{e,-1} \left(\frac{1+z}{1.02}\right)^{\frac{3}{\delta +8}}\, n_{-4}^{-\frac{1}{\delta+8}} \,\tilde{E}_{49}^{\frac{1}{\delta+8}}\,t_{15\,{\rm d}}^{-\frac{3}{\delta+8}}\,,
\end{eqnarray}
}
\noindent with $\epsilon_{e}$ being the microphysical parameter associated to the energy fraction given to accelerate electrons.   Similarly,  it is possible to define the microphysical parameter $\epsilon_{B}=U_{B}/U$ (with $U_{B}=B'^2/8\pi$) associated to the fraction of the energy given to generate and/or amplify the comoving magnetic field, which is given by
{\small
\be
B'=0.4\,\,{\rm mG}\,\left(\frac{1+z}{1.02}\right)^{\frac{3}{\delta +8}}\, \epsilon^{\frac12}_{B,-3}\,n_{-4}^{\frac{\delta+6}{2(\delta+8)}}  \,\tilde{E}^{\frac{1}{\delta+8}}_{49}\,t_{15\,{\rm d}}^{-\frac{3}{\delta+8}}\,.
\ee
}
Requiring the equality of the deceleration (eq. \ref{t_dec}) and synchrotron timescales, the cooling electron Lorentz factor becomes   
{\small
\be\label{gammac}
\gamma_{\rm c}=   2.7\times 10^8\,\, \left(\frac{1+z}{1.02}\right)^{\frac{\delta-1}{\delta +8}}\, \epsilon^{-1}_{B,-3}\,n_{-4}^{-\frac{\delta+5}{\delta+8} }\,\tilde{E}^{-\frac{3}{\delta+8}}_{49}\,t_{15\,{\rm d}}^{\frac{1-\delta}{\delta+8}}\,.
\ee
}
\noindent Comparing the acceleration and synchrotron timescales, the maximum Lorentz factor is 
{\small
\be\label{gamma_max}
\gamma_{\rm max}=8.4\times 10^8\,\phi^{1/2}_{-1}\left(\frac{1+z}{1.02}\right)^{\frac{3}{2(\delta +8)}}\,n_{-4}^{\frac14}\,\tilde{E}_{49}^{\frac{1}{2(\delta+8)}}\,t_{15\,{\rm d}}^{-\frac{3}{2(\delta+8)}}\,,
\ee
}
with $\phi$ the efficiency parameter \citep[e.g, see][]{2015ApJ...804..105F}. Given the synchrotron process with eqs. (\ref{t_dec} - \ref{gammac}),  the synchrotron spectral breaks can be written as
{\small
\bary\label{energies_break}
\epsilon_{\rm m}&\simeq& 1.5\times 10^{-3}\,{\rm GHz}\,\, \left(\frac{1+z}{1.02}\right)^{\frac{4-\delta}{\delta+8}}\epsilon^2_{e,-1}\,\epsilon_{B,-3}^{\frac12}\,n_{-4}^{\frac{\delta}{2(\delta+8)}}\,\tilde{E}^{\frac{4}{\delta+8}}_{49}\cr
&&\hspace{6.3cm}  \times\,\, t_{15\,{\rm d}}^{-\frac{12}{\delta+8}} \hspace{1cm}\cr
\epsilon_{\rm c}&\simeq&  28.9\,{\rm keV}  \left(\frac{1+z}{1.02}\right)^{\frac{\delta-4}{\delta+8}} (1+Y)^{-2}\, \epsilon_{B,-3}^{-\frac32}\,n_{-4}^{-\frac{16+3\delta}{2(\delta+8)}}\,\tilde{E}^{-\frac{4}{\delta+8}}_{49}\cr
&&\hspace{6.1cm}  \times\,\,t_{15\,{\rm d}}^{-\frac{2\delta+4}{\delta+8}}\,, 
\eary
}
where $Y$ is the Compton parameter \citep[i.e. see,][]{2016ApJ...831...22F}.   For some purposes, the spectral breaks can be described through the luminosity, which is defined as  $\tilde{L}=\tilde{E}/t$.   The maximum flux estimated through the peak spectral power is given by
{\small
\bary
F_{\rm max} &\simeq& 3\times 10^{-3}\,{\rm mJy}\,\, \left(\frac{1+z}{1.02}\right)^{\frac{8-2\delta}{\delta+8}}\,\epsilon_{B,-3}^{\frac12}\,n_{-4}^{\frac{3\delta+8}{2(\delta+8)}}\, D^{-2}_{26.5}  \,\tilde{E}^{\frac{8}{\delta+8}}_{49}\cr
&&\hspace{5.8cm}\times\,t_{15\,{\rm d}}^{\frac{3\delta}{\delta+8}}\,,
\eary
}
where $D=100$ Mpc is the luminosity distance to the source.  Using the observed synchrotron spectrum in the fast- and slow-cooling regimes with eqs. (\ref{t_dec} - \ref{gamma_max}), the synchrotron light curves in the fast-cooling regime can be written as
{\small
\begin{eqnarray}
\label{fcsyn_t}
F_{\nu,f}= \cases{ 
A_{\rm fl}\,t_{15\,{\rm d}}^{\frac{11\delta+4}{3(\delta+8)}}\, \left(\frac{\epsilon_\gamma}{\rm 6\, GHz}\right)^{\frac13},\hspace{1cm} \epsilon_\gamma<\epsilon_{\rm c}, \cr
A_{\rm fm}\,t_{15\,{\rm d}}^{\frac{2\delta-2}{\delta+8}}\, \left(\frac{\epsilon_\gamma}{\rm 1\, eV}\right)^{-\frac12},\,\hspace{1cm}  \epsilon_{\rm c}<\epsilon_\gamma<\epsilon_{\rm m}, \cr
A_{\rm fh}\,t_{15\,{\rm d}}^{\frac{2\delta+4-6p}{\delta+8}}\,\left(\frac{\epsilon_\gamma}{\rm 1\, keV} \right)^{-\frac{p}{2}},\,\hspace{0.5cm}  \epsilon_{\rm m}<\epsilon_\gamma<\epsilon_{\rm max}\,, \cr
}
\end{eqnarray}
}
with coefficients given by
{\small
\bary
A_{\rm fl}&=&2.4\times 10^{-6}\,{\rm mJy}\,\left(\frac{1+z}{1.02}\right)^{\frac{28-7\delta}{3(\delta+8)}} (1+Y)^{\frac23}\,\epsilon_{B,-3}\,n_{-4}^{\frac{20+6\delta}{3(\delta+8)}}\cr
&&\hspace{5.3cm}\times\,D_{26.5}^{-2}\, \tilde{E}_{49}^{\frac{28}{3(\delta+8)}}\cr
A_{\rm fm}&=&0.6\,{\rm mJy}\,\left(\frac{1+z}{1.02}\right)^{\frac{12-3\delta}{2(\delta+8)}} (1+Y)^{-1}\,\epsilon^{-\frac14}_{B,-3}\,n_{-4}^{\frac{3\delta}{4(\delta+8)}}\cr
&&\hspace{5.5cm}\times\, D_{26.5}^{-2}\,\tilde{E}_{49}^{\frac{6}{\delta+8}}\cr
A_{\rm fh}&=&3.3\times 10^{-9}\,{\rm mJy}\,\left(\frac{1+z}{1.02}\right)^{\frac{(4-\delta)(p+2)}{2(\delta+8)}} (1+Y)^{-1}\,\epsilon^{p-1}_{e,-1}\,\epsilon^{\frac{p-2}{4}}_{B,-3}\cr
&&\hspace{4cm}\times\, n_{-4}^{\frac{\delta(p+2)}{4(\delta+8)}}\,D_{26.5}^{-2}\,\tilde{E}_{49}^{\frac{4+2p}{\delta+8}}\,.  
\eary
}
The light curve when synchrotron emission lies in the slow-cooling regime is
{\small
\begin{eqnarray}
\label{scsyn_t}
F_{\nu,s}=\cases{
A_{\rm sl}\,t_{15\,{\rm d}}^{\frac{3\delta+4}{\delta+8}} \, \left(\frac{\epsilon_\gamma}{\rm 6\, GHz}\right)^{\frac13},\hspace{1.4cm} \epsilon_\gamma<\epsilon_{\rm m},\cr
A_{\rm sm}\,t_{15\,{\rm d}}^{\frac{3\delta-6(p-1)}{\delta+8}}\,\left(\frac{\epsilon_\gamma}{\rm 1\, eV}\right)^{-\frac{p-1}{2}},\hspace{0.3cm}     \epsilon_{\rm m}<\epsilon_\gamma<\epsilon_{\rm c},\,\,\,\,\,\cr
A_{\rm sh}\,t_{15\,{\rm d}}^{\frac{2\delta+4-6p}{\delta+8}}\,\left(\frac{\epsilon_\gamma}{\rm 1\, keV} \right)^{-\frac{p}{2}},\hspace{0.8cm} \epsilon_{\rm c}<\epsilon_\gamma\,, \cr
}
\end{eqnarray}
}
with the coefficients given by
{\small
\bary
A_{\rm sl}&=& 3.9\times 10^{-2}\,{\rm mJy}\,\left(\frac{1+z}{1.02}\right)^{\frac{20-5\delta}{3(\delta+8)}}\,\epsilon^{-\frac23}_{e,-1}\,\epsilon^\frac13_{B,-3}\,n_{-4}^{\frac{4\delta+12}{3(\delta+8)}}\,D^{-2}_{26.5}\,\cr
&&\hspace{6.3cm}\times \tilde{E}^{\frac{20}{3(\delta+8)}}_{49}\cr
A_{\rm sm}&=&3.9\times \,10^{-8}\,{\rm mJy}\,\left(\frac{1+z}{1.02}\right)^{\frac{(4-\delta)(p+3)}{2(\delta+8)}}\,\epsilon^{p-1}_{e,-1}\,\epsilon^{\frac{p+1}{4}}_{B,-3}\,n_{-4}^{\frac{16+ \delta(p+5)}{4(\delta+8)}}\cr
&&\hspace{5.7cm}\times\,D^{-2}_{26.5}\,\tilde{E}^{\frac{6+2p}{\delta+8}}_{49}\,  \cr
A_{\rm sh}&=&3.3\times 10^{-9}\,{\rm mJy}\,\left(\frac{1+z}{1.02}\right)^{\frac{(4-\delta)(p+2)}{2(\delta+8)}} (1+Y)^{-1}\,\epsilon^{p-1}_{e,-1}\,\epsilon^{\frac{p-2}{4}}_{B,-3}\cr
&&\hspace{4.2cm}\times\,n_{-4}^{\frac{\delta(p+2)}{4(\delta+8)}}\,D^{-2}_{26.5}\,\tilde{E}^{\frac{4+2p}{\delta+8}}_{49}\,.
\eary
}
It is worth noting that when $\delta=0$, the observable quantities derived in \cite{1998ApJ...497L..17S} and the light curves of the synchrotron forward-shock emission are recovered \citep[e.g., see][]{2016ApJ...831...22F}.\\ 
\subsubsection{Lateral Expansion ($\Gamma_{\rm c} \sim 1/\theta_c$ and $\beta_{\rm c}\lesssim 1$)}
In a realistic case, when  $\Gamma_{\rm c}$ drops below $\theta_c$  material begins to slow down and expands laterally.  The beaming cone of the radiation emitted broaden increasingly up to  this cone reaches our field of view \citep[$\Gamma_{\rm c}\sim \theta^{-1}_c$; ][]{2000ApJ...537..785D, 2002ApJ...570L..61G, 1999A&AS..138..491R, 2017arXiv171006421G, 1999ApJ...519L..17S}. In the lateral expansion phase, the break should occur at 
{\small
\be\label{lat_exp_br}
t_{\rm br}=22.8\,{\rm d}\,\, {\rm k}\left(\frac{1+z}{1.02}\right)\, n^{-\frac{1}{3}}_{-4}\, \tilde{E}_{49}^{\frac{1}{3}}\, \beta_{\rm c}^{-\frac{\alpha_s}{3}} \,\theta_{c,15^\circ}^{\frac{\alpha_s+6}{3}}\,,
\ee
}
with $\beta_{\rm c}\simeq\sqrt{1-\theta^2_{\rm c}}$ and  ${\rm k}$ a parameter that is added to link the maximum value of the observed flux with the jet break by means of the opening and viewing angles \citep{2002ApJ...579..699N, 2002ApJ...570L..61G}. Hereafter through this paper the value of ${\rm k}=1$ will be assumed \citep{2002ApJ...579..699N, 2017arXiv171006421G}.\\
Once the synchrotron flux reaches our field of view, the synchrotron spectral breaks and the maximum flux become
{\small
\bary\label{ener_jetbreak}
\epsilon_{\rm m}&\simeq&  9.7\times 10^{-3}\,{\rm GHz}\,\, \left(\frac{1+z}{1.02}\right)^{\frac{6-\alpha_s}{\alpha_s+6}}\, \epsilon^2_{e,-1}\,\epsilon_{B,-3}^{\frac12}\,n_{-4}^{\frac{\alpha_s-2}{2(\alpha+6)}}\,\beta_{\rm c}^{-\frac{4\alpha_s}{\alpha_s+6}}\cr
&&\hspace{5.8cm}\times \tilde{E}^{\frac{4}{\alpha_s+6}}_{49}\,t_{15\,{\rm d}}^{-\frac{12}{\alpha_s+6}} \hspace{1cm}\cr
\epsilon_{\rm c}&\simeq& 16.6 \,{\rm keV}  \left(\frac{1+z}{1.02}\right)^{\frac{\alpha_s-6}{\alpha_s+6}} (1+Y)^{-2}\, \epsilon_{B,-3}^{-\frac32}\,n_{-4}^{-\frac{(3\alpha_s+10)}{2(\alpha_s+6)}}\,\beta_{\rm c}^{\frac{4\alpha_s}{\alpha_s+6}}\cr
&&\hspace{5.4cm}\times \tilde{E}^{-\frac{4}{\alpha_s+6}}_{49}\,t_{15\,{\rm d}}^{-\frac{2\alpha_s}{\alpha_s+6}}\,, 
\eary
}
and
{\small
\bary\label{flux_jetbreak}
F_{\rm max} &\simeq& 0.1\,{\rm mJy}\,\, \left(\frac{1+z}{1.02}\right)^{\frac{12-2\alpha_s}{\alpha_s+6}}\,\epsilon_{B,-3}^{\frac12}\,n_{-4}^{\frac{2-\alpha_s}{2(\alpha_s+6)}}\, D^{-2}_{26.5}\,\beta_{\rm c}^{-\frac{8\alpha_s}{\alpha_s+6}}\cr
&&\hspace{4.2cm}\times \tilde{E}^{\frac{8}{\alpha_s+6}}_{49}\,t_{15\,{\rm d}}^{-\frac{3(2-\alpha_s)}{\alpha_s+6}}\,,\,\,\,\,\,\,\,\,\,
\eary
}
respectively.  Regarding the synchrotron spectrum, synchrotron spectral breaks and maximum flux, the synchrotron light curve in the slow-cooling regime becomes
{\small
\begin{eqnarray}
\label{scsyn_jetbreak}
F_{\nu}=\cases{
A_{\rm sl}\,t_{15\,{\rm d}}^{\frac{3\alpha_s-2}{\alpha_s+6}} \, \left(\frac{\epsilon_\gamma}{\rm 6\, GHz}\right)^{\frac13},\hspace{1.3cm} \epsilon_\gamma<\epsilon_{\rm m},\cr
A_{\rm sm}\,t_{15\,{\rm d}}^{\frac{3(\alpha_s-2p)}{\alpha_s+6}}\,\left(\frac{\epsilon_\gamma}{\rm 1\, eV}\right)^{-\frac{p-1}{2}},\hspace{0.5cm}\epsilon_{\rm m}<\epsilon_\gamma<\epsilon_{\rm c},\,\,\,\,\,\cr
A_{\rm sh}\,t_{15\,{\rm d}}^{\frac{2(\alpha_s-3p)}{\alpha_s+6}}\,\left(\frac{\epsilon_\gamma}{\rm 1\, keV} \right)^{-\frac{p}{2}},\hspace{0.8cm}\epsilon_{\rm c}<\epsilon_\gamma\,, \cr
}
\end{eqnarray}
}
with the coefficients given by
{\small
\bary
A_{\rm sl}&=& 0.8\,{\rm mJy}\,\left(\frac{1+z}{1.02}\right)^{\frac{30-5\alpha_s}{3(\alpha_s+6)}}\,\epsilon^{-\frac23}_{e,-1}\,\epsilon^\frac13_{B,-3}\,n_{-4}^{\frac{4-2\alpha_s}{3(\alpha_s+6)}}\,D^{-2}_{26.5}\,\cr
&&\hspace{4.3cm}\times \beta_{\rm c}^{-\frac{20\alpha_s}{3(\alpha_s+6)}}\, \tilde{E}^{\frac{20}{3(\alpha_s+6)}}_{49}\cr
A_{\rm sm}&=& 4.5\times 10^{-6}{\rm mJy}\,\left(\frac{1+z}{1.02}\right)^{\frac{(6-\alpha_s)(p+3)}{2(\alpha_s+6)}}\,\epsilon^{p-1}_{e,-1}\,\epsilon^{\frac{p+1}{4}}_{B,-3}\cr
&&\hspace{2cm}\times\,n_{-4}^{\frac{(\alpha_s-2)(p-3)}{4(\alpha_s+6)}}\,D^{-2}_{26.5}\, \beta_{\rm c}^{-\frac{\alpha_\alpha(6+2p)}{\alpha_s+6}}  \,\tilde{E}^{\frac{6+2p}{\alpha_s+6}}_{49}\,  \cr
A_{\rm sh}&=&2.9\times 10^{-7} \,{\rm mJy}\,\left(\frac{1+z}{1.02}\right)^{\frac{(6-\alpha_s)(p+2)}{2(\alpha+6)}} (1+Y)^{-1}\,\epsilon^{p-1}_{e,-1}\cr
&&\times\epsilon^{\frac{p-2}{4}}_{B,-3}\,n_{-4}^{\frac{p(\alpha_s-2)-2(3\alpha_s+2)}{4(\alpha_s+6)}}\,D^{-2}_{26.5}\, \beta_{\rm c}^{-\frac{\alpha_s(4+2p)}{\alpha_s+6}}\, \tilde{E}^{\frac{4+2p}{\alpha_s+6}}_{49}\,.
\,\,\,\eary
}
For $\alpha_s=0$, the observable quantities derived (eqs. \ref{ener_jetbreak} and \ref{flux_jetbreak})  and the synchrotron  light curve in the lateral expansion regime are recovered \citep[i.e., see][]{2000ApJ...537..785D, 2002ApJ...570L..61G, 1999A&AS..138..491R, 2017arXiv171006421G, 1999ApJ...519L..17S}.\\ 
\subsubsection{Non-relativistic Phase  ($\beta_{\rm c}\ll 1$)}
Once the decelerated material has swept enough ambient medium,  it will go into a non-relativistic phase.  This transition affects the evolution of the  material and in turn the synchrotron light curve.  During this phase, the kinetic equivalent energy is $E_{\rm k}\propto \beta_{\rm c}^2 R^3$ with the radius $R=\beta_{\rm c} t$, the magnetic field is $B'\propto\beta_{\rm c}$ and the minimum Lorentz factor is $\gamma_{\rm m}\propto \beta_{\rm c}^{2}$.  Taking into account that the kinetic energy is given as a power law distribution   $E_{\rm k}\propto \beta_{\rm c}^{-\alpha_s}$ (eq. \ref{beta}), then the velocity evolves as $\beta_{\rm c}\propto t^{-\frac{3}{5+\alpha_s}}$.   The synchrotron spectral breaks and maximum flux evolve as

\bary
\epsilon_{\rm m}&\propto& \gamma^2_{\rm m} B' \propto \beta_{\rm c}^5\propto t^{-\frac{15}{\alpha_s+5}}\,,\cr
\epsilon_{\rm c}&\propto& B'^{-3} t^{-2}  \propto \beta_{\rm c}^{-3}t^{-2}\propto t^{-\frac{2\alpha_s+1}{\alpha_s+5}}\,,\cr
F_{\rm max}&\propto& N_eP'_{\rm max}\propto R^3B' \propto \beta_{\rm c}^4t^3\propto t^{\frac{3(\alpha_s+1)}{\alpha_s+5}}\,.
\eary
The synchrotron light curve in the slow-cooling regime when the decelerated material lies  in the non-relativistic phase becomes
{\small
\begin{eqnarray}
\label{scsyn_tnon}
F_{\nu}\propto\cases{
t^{\frac{3\alpha_s+8}{\alpha_s+5}} \, \epsilon_\gamma^{\frac13},\hspace{1.6cm} \epsilon_\gamma<\epsilon_{\rm m},\cr
t^{\frac{6\alpha_s-15p+21}{2(\alpha_s+5)}}\,\epsilon_\gamma^{-\frac{p-1}{2}},  \hspace{0.4cm}   \epsilon_{\rm m}<\epsilon_\gamma<\epsilon_{\rm c},\,\,\,\,\,\cr
t^{\frac{4\alpha_s-15p+20}{2(\alpha_s+5)}}\,\epsilon_\gamma^{-\frac{p}{2}},\,\hspace{0.6cm}\epsilon_{\rm c}<\epsilon_\gamma\,. \cr
}
\end{eqnarray}
}
It is worth noting that when $\alpha_s=0$, the observable quantities derived  and the synchrotron  light curve in the non-relativistic regime are recovered \citep[i.e., see][]{1999ApJ...519L.155D, 2003MNRAS.341..263H, 2000ApJ...538..187L, 1999MNRAS.309..513H, 1997MNRAS.288L..51W}.\\ 
\subsection{Description of the observable quantities}
\subsubsection{X-ray, optical and radio light curves}
Figures \ref{lc1} and \ref{lc2} display the resulting X-ray, optical and radio light curves of  the synchrotron forward-shock emission generated by a decelerated shock-breakout material for several parameter values.   The light curves are presented for three electromagnetic bands: X-ray at 1 keV (upper panel), optical at 1 eV (medium panel) and radio (lower panel) at 6 GHz, with the parameter values in the ranges of $10^{-6}\leq n\leq 1\,{\rm cm^{-3}}$,  $10^{-4} \leq\epsilon_B\leq 10^{-1}$, $2.2\leq p \leq 3.6$ and $2.0\leq \alpha_s \leq3.5$, for the opening angle $\theta_c=15^\circ$,  the fiducial kinetic energy of $\tilde{E}=10^{50}$ erg, the microphysical parameter given to accelerate electrons $\epsilon_{e}=0.1$ and a source located at $D=100\, {\rm Mpc}$.\\
The left- and right-hand panels in Figure \ref{lc1} display the light curves considering the magnetic microphysical parameter $\epsilon_B=10^{-2}$ and  density $n=10^{-2}\,\,{\rm cm^{-3}}$, respectively, for $p=2.2$,  $\alpha_s=3.0$, $D=100\, {\rm Mpc}$, $\theta_{\rm c}=15^\circ$, $\tilde{E}=10^{50}$ erg and $\epsilon_{e}=0.1$.  Figure \ref{lc2} shows  the light curves considering the power indexes  $\alpha_s=3.0$ (left-hand panels) and  $p=2.2$ (right-hand panels), respectively, for $\epsilon_B=10^{-2}$,  $n=10^{-2}\,\,{\rm cm^{-3}}$, $D=100\, {\rm Mpc}$,  $\theta_c=15^\circ$, $\tilde{E}=10^{50}$ erg and $\epsilon_{e}=0.1$.\\
Figure \ref{lc1} shows that X-ray, optical and radio light curves have similar behaviors depending on which power-law segment of the spectrum are evolving. The left-hand panels show that for $n=1\, {\rm cm^{-3}}$, light curves do not present peaks, and as density decreases, the peak in the light curves are more notable. The right-hand panels display that radio flux is evolving in the same power law for $10^{-4}\leq \epsilon_B\leq 10^{-1}\,{\rm cm^{-3}}$ whereas X-ray and optical fluxes evolve in different power-law segments of the synchrotron spectrum. For instance,  the optical flux for $\epsilon_B=0.1$ changes  the power-law segment during the jet break. After the jet break, it evolves in the second power law while for other values of $\epsilon_B$ it keeps evolving in the first one.\\
Figure \ref{lc2} shows that the fluxes are strongly dependent on the values of $p$ and $\alpha_s$.  The right-hand panels show that at early times  fluxes, in general, are dominated by those generated with small values of $\alpha_s$  whereas at later times are dominated by those with larger values.  The left-hand panels show that as the electron power-law index increases, fluxes, in general, decrease, with the exception of the radio flux  at $\epsilon_\gamma=$ 6 GHz for p=2.2.  This is due to the fact that when $\epsilon_{\rm m}$  scales as $\left(\frac{p-1}{p-2}\right)^2$, then radio flux for p=2.6, 3.2 and 3.5 is in the regime $\epsilon_\gamma< \epsilon_{\rm m}$ which in turn does not depend on $p$ (as shown in these panels for early times) while for p=2.2 is in the regime $\epsilon_{\rm m}< \epsilon_\gamma$. With the passage of time, $\epsilon_{\rm m}$ becomes less than 6 GHz  and then the radio flux changes to the regime $\epsilon_{\rm m}> \epsilon_\gamma$, firstly for p=2.6, then for p=3.2 and finally for p=3.6, as shown in these panels for later times. While radio and optical fluxes peak during the first  hundred days,  X-ray flux peaks depending the value of $\alpha_s$.  Furthermore, the right-hand panels show that as  $\alpha_s$ increases, X-ray fluxes peak earlier.\\
\vspace{1.5cm}
\subsubsection{Evolution of the spectral breaks}
In the previous subsection the X-ray, optical and radio light curves were illustrated for deceleration timescales ranging from hours to one thousand days.  In this subsection we illustrate the evolution of the synchrotron spectral breaks  during the first seconds. In this case,  the deceleration time is around  a few seconds and the bulk Lorentz factor close to one hundred.   The ranges of values for the microphysical parameters and ISM used in this analysis are those that allow the cooling spectral breaks to be observed in the energy range covered by Fermi-GBM, Swift-BAT and INTEGRAL.\\ 
Figure \ref{Epeak} shows the evolution of synchrotron cooling and characteristic energies  during the first 10 s {for typical values  of the magnetic microphysical parameter  ({\small $10^{-2}<\epsilon_B<10^{-1}$}; upper panel), the ISM ({\small $10^{-2}<n<1\,{\rm cm^{-3}}$}; medium panel) and the fiducial energy  ({\small $10^{49} <\tilde{E}<10^{51}\, {\rm erg}$}; lower panel).  The left-hand panels display that $\epsilon_c$ evolves in the $\gamma$-ray band as $t^{-1.09}$ ($\alpha_s=3.0$) or $t^{-1.14}$ ($\alpha_s=4.0$)  and  the right-hand panel shows that  $\epsilon_m$ evolves in the IR - optical band as $t^{-0.92}$ ($\alpha_s=3.0$) or $t^{-0.85}$ ($\alpha_s=4.0$). While the evolution of the cooling spectral break can be potentially detected by instruments such as Fermi-GBM and Swift-BAT, the characteristic spectral break could be detected by Swift-UVOT, 1-meter Swope telescope and other optical telescopes.   It is worth noting that our results are different to the evolution of synchrotron cooling $t^{-0.5}$ and characteristic $t^{-1.5}$ energies derived when the relativistic jet is decelerated by the ISM \citep[e.g. see,][]{1998ApJ...497L..17S}.\\
\\
\section{Relativistic off-axis Jet}
We consider that a relativistic jet producing the long-lived afterglow emission  is  launched from a binary NS system.  Additionally, we assume the relativistic jet is not aligned with the observer's line of sight \citep{2017ApJ...848L..34M, 2017arXiv171005905I, 2017arXiv171203237L, 2017arXiv171008514F, 2017arXiv171006421G}.  In order to obtain the quantities for the off-axis emission,  the quantities derived in \cite{1998ApJ...497L..17S} have to be modified using different boosts.  For the off-axis case,  given the evolution of the minimum electron Lorentz factor $\gamma_{\rm m, off}\propto \Gamma_{\rm j}$, the magnetic field $B'\propto \Gamma_{\rm j}$, the cooling electron Lorentz factor $\gamma_{\rm c, off}\propto \delta_D^{-1}\,\Gamma_{\rm j}^{-2}\,t^{-1}$, the number of radiating electrons $N_{\rm e, off}\propto R^3_{\rm off}\propto (\delta_D \Gamma_{\rm j} t)^3$ \citep{2003ApJ...592.1002S},  the solid angle  $\Omega_{\rm off}\propto\delta^{-2}_D$   \citep{1986rpa..book.....R},  the maximum power $P_{\rm \nu_m, off}\propto \delta_D B'$,  the synchrotron energy breaks and the maximum synchrotron flux  evolve as
\bary\label{quant_off}
\epsilon_{\rm m, off}&\propto& \delta_D B' \gamma_{\rm m, off}^2 \propto   \delta_D\Gamma_{\rm j}^3\,, \cr
\epsilon_{\rm c, off}&\propto&   \delta_D B'  \gamma_{\rm c, off}^2   \propto  \Gamma_{\rm j}^{-3}\delta_D^{-1}t^{-2}\,,\cr
F_{\rm \nu,max, off}&\propto&\frac{N_{\rm e, off}\,P_{\rm \nu_m, off}}{\Omega_{\rm off}}\propto \delta_D^3 N_{\rm e, off} B'\propto \delta_D^6 \Gamma_{\rm j}^{4}t^3,
\eary
respectively. Therefore, the flux density at a given energy evolves as $F_{\rm \nu,off}\propto \delta_D^\frac{17}{3}\Gamma_{\rm j}^3 t^3$ for $\epsilon_{\rm \gamma}\leq \epsilon_{\rm m}$,  $\propto\delta_D^\frac{p+11}{2}\Gamma_{\rm j}^\frac{3p+5}{2}t^3$ for $ \epsilon_{\rm m} \leq \epsilon_{\rm \gamma}\leq \epsilon_{\rm c} $ and $\propto\delta_D^\frac{p+10}{2}\Gamma_{\rm j}^\frac{3p+2}{2}t^2$ for $ \epsilon_{\rm c} \leq \epsilon_{\rm \gamma}$, where  the Doppler factor is  $\delta_D=\delta_D (\Delta \theta) \equiv\frac{1}{\Gamma_{\rm j}(1-\beta_{\rm j} \Delta \theta)}\approx\frac{2\Gamma_{\rm j}}{1+\Gamma_{\rm j}^2\Delta\theta^2}$ for $\Gamma_{\rm j}\gg 1$ and $\Delta \theta \ll1 $  with $\Delta\theta= \theta_{\rm obs} - \theta_j$. The value of the flux density for  $ \epsilon_{\rm m} \leq \epsilon_{\rm \gamma}\leq \epsilon_{\rm c}$ agrees with the flux reported in \cite{2002MNRAS.332..945R} and \cite{2018arXiv180109712N}.  It is worth noting that when $\Delta\theta=0^\circ$ and $F_{\rm \nu,max, off}$ is divided by $\Omega=4\pi\delta^2_D$ \citep{2018MNRAS.481.2581L}, the relations derived in \cite{1998ApJ...497L..17S}  are recovered.\\
\\
In order to find the evolution of the bulk Lorentz factor ($\Gamma_{\rm j}$) for a decelerated off-axis jet, the kinetic equivalent energy calculated through the Blandford-McKee condition $E_{\rm k, j}=\frac{4}{3}\pi m_p\, n\, R^3_{\rm off}\,\Gamma_{\rm j}^2$ is obtained as follows.  Taking into account  the deceleration radius viewed off-axis $R_{\rm off}=\frac{\delta_D\Gamma_{\rm j}}{(1+z)}t$\footnote{For $\Delta \theta=0$,  the quantities become $\delta_D\simeq \Gamma_{\rm j}$ and  $R_{\rm off}\simeq R$, and the equivalent kinetic on-axis energy is recovered \citep{2003ApJ...592.1002S}.} and the Doppler boost $\delta_D\simeq\frac{2}{\Gamma_{\rm j}  \Delta \theta^2}$ for $\Gamma_{\rm j}^2\Delta \theta^2\gg1$ with the energy limited to the opening angle $E_{\rm k,j}=\frac{\tilde{E}}{1-\cos \theta_j}\simeq \frac{2\tilde{E}}{\theta_j^2}$ \citep{2017arXiv171006421G}, the kinetic energy becomes $\tilde{E}=\frac{16}{3}\pi m_p\,(1+z)^{-3} n\,\theta_j^2\, t^3\, \Gamma_{\rm j}^2\Delta \theta^{-6}$\footnote{Other way to derive the equivalent kinetic energy is given as follows. The corresponding kinetic energy viewed off-axis for $\theta_{\rm obs}> 2\theta_j$ is given by $E_{\rm k,j}= \left( \frac{\delta_D(0)}{\delta_D(\Delta \theta)} \right)^3 E_{\rm k, on}$ \citep{2017arXiv171005905I}.   Considering the Blandford-McKee condition,   the equivalent kinetic off-axis energy is given by $E_{\rm k,j}=\frac{32}{3}\pi m_p\,(1+z)^{-3} n\, t^3\, \Gamma_{\rm j}^8 (1+ \Gamma_{\rm j}^2\Delta \theta^2)^{-3}$.    Taking into account that energy is limited to the opening angle $E_{\rm k,j}\simeq \frac{2\tilde{E}}{\theta_j^2}$ \citep{2017arXiv171006421G}, the kinetic energy becomes $\tilde{E}=\frac{16}{3}\pi m_p\,(1+z)^{-3} n\,\theta_j^2\, t^3\, \Gamma_{\rm j}^2\Delta \theta^{-6}$ for $\Gamma_{\rm j}^2\Delta \theta^2\gg1$.}.    
In this case, the bulk Lorentz  factor evolves as
{\small
\be\label{Gamma}
\Gamma_{\rm j}=8.3\,(1+z)^{\frac{3}{2}}  \,n^{-\frac{1}{2}}_{-4}\,\tilde{E}^{\frac{1}{2}}_{50}\,\theta_{j,5^\circ}^{-1}\,\Delta\theta_{15^\circ}^3\,t^{-\frac{3}{2}}_{\rm 100d}\,.
\ee
}
Using the bulk Lorentz factor (eq. \ref{Gamma}) and eqs. (\ref{quant_off}), we derive the relevant quantities of forward-shock synchrotron emission radiated in an off-axis jet. The minimum and cooling electron Lorentz factors are  given by
{\small
\begin{eqnarray}\label{gammas}
\gamma_{\rm m,off}&=& 212.9\, \epsilon_{e,-1} (1+z)^{\frac{3}{2}}\, n_{-4}^{-\frac{1}{2}} \,\tilde{E}_{50}^{\frac{1}{2}}\,\Delta\theta_{15^\circ}^{3}\,\theta_{j,5^\circ}^{-1}\,t_{100\,{\rm d}}^{-\frac{3}{2}}\,,\cr
\gamma_{\rm c,off}&=& 9.1\times 10^6\,\, (1+z)^{-\frac{1}{2}}\,(1+Y)^{-1} \epsilon^{-1}_{B,-4}\,n_{-4}^{-\frac{1}{2} }\,\tilde{E}^{-\frac{1}{2}}_{50}\,\Delta\theta_{15^\circ}^{-1}\,\theta_{j,5^\circ}\,\cr
&&\hspace{6.3cm}\times\,t_{100\,{\rm d}}^{\frac{1}{2}}\,,
\end{eqnarray}
}
which correspond to a comoving magnetic field given by
{\small
\be
B'=1.9\times 10^{-2}\,\,{\rm mG}\,(1+z)^{\frac{3}{2}}\, \epsilon^{\frac12}_{B,-4}\, \,\tilde{E}^{\frac{1}{2}}_{50}\,\Delta\theta_{15^\circ}^{3}\,\theta_{j,5^\circ}^{-1}\,t_{100\,{\rm d}}^{-\frac{3}{2}}\,.
\ee
}
The synchrotron spectral breaks and the maximum flux can be written as
{\small
\bary\label{energies_break}
\epsilon_{\rm m,off}&\simeq& 1.5\times 10^{-2}\,{\rm GHz}\,\, (1+z)^2\epsilon^2_{e,-1}\,\epsilon_{B,-4}^{\frac12}\,n_{-4}^{-\frac{1}{2}}\,\tilde{E}_{50}\,\Delta\theta_{15^\circ}^{-4}\,\theta_{j,5^\circ}^{-2}\,\cr
&&\hspace{6.6cm}\times   t_{100\,{\rm d}}^{-3}, \hspace{1cm}\,\cr
\epsilon_{\rm c,off}&\simeq&  4.1\,{\rm keV}  (1+z)^{-2} (1+Y)^{-2}\, \epsilon_{B,-4}^{-\frac32}\,n_{-4}^{-\frac{1}{2}}\,\tilde{E}^{-1}_{50}\Delta\theta_{15^\circ}^{-4}\theta_{j,5^\circ}^{2}\,\cr
&&\hspace{6.3cm}\times\,t_{100\,{\rm d}}\,,\cr 
F_{\rm max,off} &\simeq& 5.6\times 10^3\,{\rm mJy}\,\, (1+z)^{-4}\,\epsilon_{B,-4}^{\frac12}\,n_{-4}^{\frac{5}{2}}\, D^{-2}_{26.5}\,\tilde{E}^{-1}_{50}\,\Delta\theta_{15^\circ}^{-18}\cr
&&\hspace{5.6cm}\times\,\theta_{j,5^\circ}^{2}\, t_{100\,{\rm d}}^{6}\,.
\eary
}
Using the observed synchrotron spectrum in the slow-cooling regime with eqs. (\ref{energies_break}), the synchrotron light curves can be written  as
{\small
\begin{eqnarray}
\label{scsyn_t}
F_{\nu,off}=\cases{
A_{\rm sl}\,t_{100\,{\rm d}}^7 \, \left(\frac{\epsilon_\gamma}{\rm 6\, GHz}\right)^{\frac13},\hspace{1.3cm} \epsilon_\gamma<\epsilon_{\rm m},\cr
A_{\rm sm}\,t_{100\,{\rm d}}^{\frac{3(5-p)}{2}}\,\left(\frac{\epsilon_\gamma}{\rm 1\, eV}\right)^{-\frac{p-1}{2}},\epsilon_{\rm m}<\epsilon_\gamma<\epsilon_{\rm c},\,\,\,\,\,\cr
A_{\rm sh}\,t_{100\,{\rm d}}^{\frac{16-3p}{2}}\,\left(\frac{\epsilon_\gamma}{\rm 1\, keV} \right)^{-\frac{p}{2}},\,\epsilon_{\rm c}<\epsilon_\gamma\,, \cr
}
\end{eqnarray}
}
with the coefficients given by
{\small
\bary
A_{\rm sl}&=& 1.8\times 10^{4}\,{\rm mJy}\,(1+z)^{-\frac{14}{3}}\,\epsilon^{-\frac23}_{e,-1}\,\epsilon^\frac13_{B,-4}\,n_{-4}^{\frac{8}{3}}\,D^{-2}_{26.5}\, \tilde{E}^{-\frac{4}{3}}_{50}\,\cr
&&\hspace{5.2cm}\times\, \Delta\theta_{15^\circ}^{-\frac{58}{3}}\,\theta_{j,5^\circ}^\frac83\,, \cr
A_{\rm sm}&=&1.5\times 10^{-2}\,{\rm mJy}\,(1+z)^{p-5}\,\epsilon^{p-1}_{e,-1}\,n_{-4}^{-\frac{(p-11)}{4}}\,\epsilon^{\frac{p+1}{4}}_{B,-4}\,D^{-2}_{26.5}\,\tilde{E}^{\frac{p-3}{2}}_{50}\cr
&&\hspace{4.5cm}\times\, \Delta\theta_{15^\circ}^{2(p-10)}\, \theta_{j,5^\circ}^{(1-p)}\,, \cr
A_{\rm fh}&=&3.5\times 10^{-4}\,{\rm mJy}\,(1+z)^{p-6} (1+Y)^{-1}\,\epsilon^{p-1}_{e,-1}\,\epsilon^{\frac{p-2}{4}}_{B,-4}\,D_{26.5}^{-2}\cr
&&\hspace{2.1cm}\times\, n_{-4}^{-\frac{(p-10)}{4}}\,\tilde{E}_{50}^{\frac{p-4}{2}}\, \Delta\theta_{15^\circ}^{2(p-11)}\,\theta_{j,5^\circ}^{(2-p)}\,.  
\eary
}
\subsection{Lateral expansion}
In this case,  the beaming cone of the radiation emitted off-axis,  $\Delta \theta$, broaden increasingly until this cone reaches our field of view \citep[$\Gamma_{\rm j}\sim \Delta \theta^{-1}$; ][]{2002ApJ...570L..61G,  2017arXiv171006421G}. In the lateral expansion phase, the break in the density flux should occur around 
{\small
\be\label{lat_exp_jt}
t_{\rm br,off}=205.6\,{\rm d}\,\,{\rm k} \,\left(\frac{1+z}{1.02}\right)\, n^{-\frac{1}{3}}_{-4}\, \tilde{E}_{50}^{\frac{1}{3}}\,  \Delta \theta_{15^\circ}^2\,,
\ee
}
In this  phase, the synchrotron spectral breaks and the maximum synchrotron flux  are
{\small
\bary\label{e_break}
\epsilon_{\rm m,off}&\simeq& 7.6\times 10^{-3}\,{\rm GHz}\, (1+z)\,\epsilon^2_{e,-1}\,\epsilon_{B,-4}^{\frac12}\,n_{-4}^{-\frac{1}{6}}\,  \tilde{E}^{\frac{2}{3}}_{50}\, t_{100\,{\rm d}}^{-2}\,,\,\,\, \cr
\epsilon_{\rm c,off}&\simeq&  2.6\,{\rm keV} \, (1+z)^{-1}    (1+Y)^{-2}\, \epsilon_{B,-4}^{-\frac32}\,n_{-4}^{-\frac{5}{6}}  \,\tilde{E}^{-\frac{2}{3}}_{50}\,,\cr
F_{\rm max,off} &\simeq& 2.5\times 10^3\,{\rm mJy}\, (1+z)^2\,\epsilon_{B,-4}^{\frac12}\,n_{-4}^{\frac{1}{6}}\,D^{-2}_{26.5}  \,\tilde{E}^{\frac{4}{3}}_{50}\,t_{100\,{\rm d}}^{-1}.\cr
&&\hspace{5cm}
\eary
}
Given the synchrotron spectrum, the synchrotron spectral breaks and the maximum flux (eq. \ref{e_break}) during this phase, the light curve when in the slow-cooling regime is
{\small
\begin{eqnarray}
\label{scsyn_t}
F_{\nu,,off}=\cases{
A_{\rm sl}\,t_{100\,{\rm d}}^{-\frac{1}{3}} \, \left(\frac{\epsilon_\gamma}{\rm 6\, GHz}\right)^{\frac13},\hspace{1.3cm} \epsilon_\gamma<\epsilon_{\rm m},\cr
A_{\rm sm}\,t_{100\,{\rm d}}^{-p}\,\left(\frac{\epsilon_\gamma}{\rm 1\, eV}\right)^{-\frac{p-1}{2}},\epsilon_{\rm m}<\epsilon_\gamma<\epsilon_{\rm c},\,\,\,\,\,\cr
A_{\rm sh}\,t_{100\,{\rm d}}^{-p}\,\left(\frac{\epsilon_\gamma}{\rm 1\, keV} \right)^{-\frac{p}{2}},\,\epsilon_{\rm c}<\epsilon_\gamma\,, \cr
}
\end{eqnarray}
}
with the coefficients given by
{\small
\bary
A_{\rm sl}&=& 9.9\times 10^2\,{\rm mJy}\,(1+z)^{\frac{5}{3}}\,\epsilon^{-\frac23}_{e,-1}\,\epsilon^\frac13_{B,-4}\,n_{-4}^{\frac{2}{9}}\,D^{-2}_{26.5}\, \tilde{E}^{\frac{10}{9}}_{50}\,,     \cr
A_{\rm sm}&=&4.4\times \,10^{-4}\,{\rm mJy}\,(1+z)^{\frac{(p+3)}{2}}\,\epsilon^{p-1}_{e,-1}\,\epsilon^{\frac{p+1}{4}}_{B,-4}\,n_{-4}^{-\frac{(p-3)}{12}}\cr
&&\hspace{4.9cm}\times\,D^{-2}_{26.5}\,\tilde{E}^{\frac{(p+3)}{3}}_{50}\,, \cr
A_{\rm sh}&=&1.7\times 10^{-5}\,{\rm mJy}\,(1+z)^{\frac{(p+2)}{2}} (1+Y)^{-1}\,\epsilon^{p-1}_{e,-1}\,\epsilon^{\frac{p-2}{4}}_{B,-4}\cr
&&\hspace{3.4cm}\times\,n_{-4}^{-\frac{(p+2)}{12}}\,D^{-2}_{26.5}\,\tilde{E}^{\frac{(p+2)}{3}}_{50} \,.
\eary
}
The observed fluxes agree with those reported in \cite{2017arXiv171006421G}.
\section{Case of application: GRB 170817A}
\subsection{Multiwavelength Afterglow Observations}
\paragraph{X-ray data.}  During the first week after the GBM trigger,   an X-ray campaign was performed without any detections, although  upper limits were placed \citep[i.e. see][]{2017ApJ...848L..20M, troja2017a}.  From the 9th up to 256th day after the GW trigger, X-ray fluxes have been detected  by the Chandra and XMM-Newton satellites \citep{troja2017a, 2018arXiv180103531M, 2018arXiv180502870A, 2018arXiv180106164D, 2017ATel11037....1M, 2018ATel11242....1H}.\\
\paragraph{Optical data}  A thermal electromagnetic counterpart, in the infrared and optical bands,  was detected at $\sim$ 11 hours after the GBM trigger \citep[see for e.g.][and references therein]{2017arXiv171005452C}, and after the first detection multiple infrared/optical telescopes followed this event \citep{2017arXiv171005841S}. Later, weak optical fluxes  \citep[$m=26.54\pm 0.14$;][]{2018arXiv180102669L} mag and \citep[$m=26.90\pm 0.25$;][]{2018arXiv180103531M}  were reported by Hubble Space Telescope (HST) at $\sim$ 110 and 137 days after the GW trigger.\\
\paragraph{Radio data.}  The VLA and  ALMA  reported radio upper limits during the first two weeks after the GW event.  On the sixteenth day after the GW trigger, and for more than seven months, the radio flux at 3 and 6 GHz was reported by Very Large Array \citep[VLA; ][]{2041-8205-848-2-L12, 2017arXiv171111573M, 2018ApJ...858L..15D, 2017Natur.547..425T}.\\
\subsection{Modelling the afterglow  emission}
Our afterglow model presented through  the shock-breakout material and the off-axis jet  is dependent on a set of 8 parameters, $\Xi_{\rm fow, b}$ = \{ $\tilde{E}$, n, p, $\theta_j$, $\Delta\theta$, $\varepsilon_B$, $\varepsilon_e$, $\alpha_s$\}\footnote{Due to the merger shock-breakout material is viewed on-axis, we have considered that the opening angle could be approximated as $\theta_c\approx \Delta \theta$}.  To find an adequate set of values inside the parameter space we utilize the Markov-Chain Monte Carlo (MCMC) method, a Bayesian statistical technique that allows us to find best-fit values through a sampling process.  By using these parameters, we determine a suitable prior distribution, to be utilized alongside a normal likelihood upon which an eight parameter $\sigma$ is introduced, that allows us to generate samples for the posterior distributions of our on-axis model. We utilize the No-U-Turn Sampler (NUTS) from the PyMC3 python distribution \citep{peerj-cs.55} to generate a total 21000 samples, of which 7000 are utilized for tuning and subsequently discarded. In all simulations we have employed a set of normal, continuous, distributions for our priors. With this specific choice we can simulate a more unbiased and uninformative set of priors.\\
Output  is presented in the Figures \ref{fig3:param_late_3R},  \ref{fig3:param_late_6R} and \ref{fig3:param_late_X}, alongside Table \ref{table5:param_late}. These figures are Corner Plots \citep{2016JOSS....1...24F}, a specific plot where the diagonal is a one-dimensional kernel density estimation (KDE) describing the posterior distribution and the lower triangle is the bi-dimensional KDE. The best fit value of each parameter is shown in green color. Table \ref{table5:param_late} reports our chosen quantiles (0.15,0.5,0.85) retrieved from inference.\\
In accordance with the values obtained after describing the multiwavelength light curves and the SEDs at 15 $\pm$ 2, 110 $\pm$ 5 and  145$\pm$ 20 days of GRB170817A, we found that i) for the shock-breakout material  the bulk Lorentz factor becomes $\Gamma_{\rm c}\simeq 3.1 \left(t/15\,d\right)^{-0.24}$,  the equivalent kinetic energy and the observed electromagnetic energy $E_{\rm k, c}=3\times 10^{49}\,{\rm erg}\,\, \left(\beta_{\rm c}\Gamma_{\rm c}\right)^{-4.3}\simeq 3.8\times 10^{47}\,{\rm erg}$ and $E_{\rm obs, \gamma}=5.4\times 10^{46}\,{\rm erg}$, respectively, which corresponds to an efficiency of $\sim 16\%$  and ii) for the relativistic off-axis jet the bulk Lorentz factor becomes $\Gamma_{\rm j}\simeq 4.6 \left(t/140\,d\right)^{-\frac32}$  the equivalent kinetic energy and the released electromagnetic energy are $E_{\rm k,j}\simeq  \frac{2\tilde{E}}{\theta^2_j}\simeq 7.01\times 10^{51}\,{\rm erg}\,$ and $E_{\rm \gamma}=5.4\times 10^{46}\,{\rm erg} (1+\Delta\theta^2\Gamma_{\rm j}^2)^{3}\simeq  4.8\,\times10^{51}\,{\rm erg}  $, respectively, which corresponds to an efficiency of $\sim 11\%$.  For the shock-breakout material, the flux evolves as  $F_\nu\propto t^{0.64}$ for  $\epsilon_{\rm m}<\epsilon_\gamma<\epsilon_{\rm c}$ and  $F_\nu\propto t^{0.09}$ for $\epsilon_{\rm c}<\epsilon_\gamma$. The cooling spectral break $\epsilon_{\rm c}\sim11.2$ keV is above the X-ray band and the characteristic break $\epsilon_{\rm m}=0.02$ GHz is below the radio band at 15 days. The X-ray, optical and radio fluxes peak at $\sim$ 30 days, and later they evolve as $F_\nu\propto t^{-0.76}$. At $\sim$ 500 s, the X-ray flux will vary from $\propto t^{-0.76}$ to  $\propto t^{-1.03}$ whereas  the optical and radio fluxes will continue evolving as $F_\nu\propto t^{-0.76}$.   During the non-relativistic phase ($\beta_{\rm c}\ll 1$), the flux will evolve in accordance with the synchrotron spectrum given in eq.~(\ref{scsyn_tnon}).   For the relativistic off-axis jet,  the cooling spectral break, $\epsilon_{\rm c}\sim3.5$ keV, is above the X-ray band and the characteristic break, $\epsilon_{\rm m}=0.03$ GHz, is below the radio band at 100 days. During this period, the observed flux increases as $F_\nu\propto t^{4.2}$ (eq. \ref{scsyn_t}), as predicted in \cite{2018arXiv180109712N}.  The X-ray, optical and radio fluxes peak at $\sim$ 140 days, and later they evolve as $F_\nu\propto t^{-2.2}$. \\
\\
On the other hand, taking into consideration the GBM data during the first seconds, \cite{2018arXiv180207328V} showed the evolution of the main peak ($E_{\rm peak}$) as a function of time, and the luminosity as a function of $E_{\rm peak}$. Using  simple power laws in both cases $E_{\rm peak} \propto (t-t_{\rm 0})^k$  and $L\propto E^q_{\rm peak}$, they obtained the best-fit values of  $k=-0.97\pm 0.35$ for $t_0=-0.15\pm 0.04\, {\rm s}$  and $q=0.90\pm0.10$.   The values of $k$ and $q$ are consistent with the evolution of the cooling spectral break (eq. \ref{energies_break}) $\epsilon_c\propto t^{-(1.01 - 1.13)}$ and  $\tilde{L} \propto \epsilon^{\frac{\delta+8}{2\delta+4}}_c=\epsilon^{0.88-0.93}_c$ for $3.0<\alpha_s<4.0$.  In order to find the values of parameters that reproduce this evolution, Figure \ref{Epeak_data} is presented.  This figure exhibits   the energy peak as a function of time since GW trigger. The red solid line is the fitted simple power law $E_{\rm peak}=A_0 (t-t_0)^{\alpha}$ with  $A_0=30.2\pm 7.97\, {\rm keV}$, $t_0=-0.15\pm0.04\, {\rm s}$, $\alpha=-0.97\pm0.35$, $\chi^2/{\rm ndf}=0.44/4$. Dashed, dotted and dashed-dotted lines represent the cooling spectral break of our theoretical model for $\alpha_s=2.3$ and $n= 0.6\,{\rm cm^{-3}}$ (green line), $\alpha_s=3.2$ and $n= 4\,{\rm cm^{-3}}$ (gold line) and $\alpha_s=4.2$ and $n= 12\,{\rm cm^{-3}}$ (blue line), respectively. The values used are  $\tilde{E}=3\times10^{49}\,\,{\rm erg}$, $\epsilon_B=10^{-1}$, $\epsilon_e=10^{-1}$ and $p=2.2$. In all the cases, the cooling spectral break  can successfully  describe the early evolution of $E_{\rm peak}$ for $1.2<\alpha_s<4.2$ and $0.6<n<10\,{\rm cm^{-3}}$. If this is the case, the bulk Lorentz factor begins to change from $\sim$ 12 at a few seconds to $\sim$ 3 around one hundred  days.    If the cooling spectral break evolution enters the X-ray band, then the density of the circumburst medium must be much higher than the one found to explain the afterglow at tens of days.  The high and low values of the ISM found at early and late times, respectively, could be explain through a stratified medium around the merger as discussed in other sGRBs  \citep{2018ApJ...853L..13W, 2009arXiv0904.1768P, 2011ApJ...735...23N}.\\
\\
\section{Discussion and Conclusion}
In the framework of the binary NS system,  we have derived the dynamics of the forward shock and the synchrotron light curves from the outermost (shock-breakout) material and the relativistic off-axis jet.  The resulting  equivalent kinetic energy for the shock-breakout material  is given by $\tilde{E} \left(\beta_{\rm c} \Gamma_{\rm c}\right)^{-\delta}$  and for the relativistic jet is $2\tilde{E}/\theta^2_j$. We have analyzed the case in which the shock-breakout material and the relativistic jet are decelerated by a homogeneous medium and evolve in the fully adiabatic regime.    The main differences between both ejected materials lie in the fact that the jet is moving at ultra-relativistic velocities, is  narrowly collimated  and is observed from a viewing angle. Taking into account the velocity regime,  the values obtained of $\alpha_s$ for  the mildly-relativistic shock-breakout material ($\alpha_s=2.3$)  agrees with the description presented by \cite{2001ApJ...551..946T} and the works previously described in \cite{2015MNRAS.450.1430H} and \cite{2014MNRAS.437L...6K}.  Considering the observed times of the lateral expansion for the shock-breakout material  (eq. \ref{lat_exp_br})  and  the relativistic off-axis jet (eq. \ref{lat_exp_jt}), the observed flux of shock-breakout material peaks first due to the emission is released on-axis from a  material described by a power-law velocity distribution.   It is worth mentioning that for $\delta=0$, the observable quantities  and light curves derived in \cite{1998ApJ...497L..17S, 2016ApJ...831...22F,  1999ApJ...519L.155D, 2003MNRAS.341..263H, 2000ApJ...537..785D, 2002ApJ...570L..61G, 1999A&AS..138..491R} are recovered.    In addition, we have analyzed  the shock-breakout material considering the values in the typical ranges: the medium density  ($10^{-6}\leq n \leq 1\,{\rm cm^{-3}}$), magnetic microphysical parameter ($10^{-4}\leq \epsilon_B \leq 10^{-1}$),   power indices ($2.2 \leq p \leq 3.6$) and  ($1.5\leq \alpha_s\leq 3.5$) for a fiducial kinetic energy ($\tilde{E} = 10^{49}$ - $10^{50}\,{\rm erg}$),  microphysical parameter ($\epsilon_e = 10^{-1}$) for an event located at a luminosity distance of $D= 100\, {\rm Mpc}$.\\
\\
For the shock-breakout material we have found that the cooling spectral break evolves in the $\gamma$-ray bands and the characteristic break in the IR - optical bands. The cooling (characteristic) spectral break evolves as  $\epsilon_c\propto t^{-(1.0 - 1.1)}$ ($\epsilon_m \propto t^{-(0.9 - 1.0)}$), instead of the typical evolution  $\epsilon_c\propto t^{-0.5}$ ($\epsilon_m\propto t^{-1.5}$) suggested by the decelerated jet \citep{1998ApJ...497L..17S}.  The analysis of the early spectral evolution of the tails as suggested by some authors \citep[e.g. see][]{1999ApJ...524L..47G, 2012ApJ...751...33F, 2017ApJ...848...15F}  could illustrate whether the evolution of $E_{\rm peak}$ as early observed in  $\gamma$-ray and optical bands could be generated by external shocks during the prompt phase. In addition, this analysis could reveal the type of scenario (e.g. internal or external shocks),  circumburst medium (e.g homogeneous or stratified), the regime (e.g. adiabatic or radiative) and the geometry of material that has been decelerated.\\ 
Considering the multiwavelength campaign dedicated to follow-up the electromagnetic counterpart of GW170817 \citep{2017arXiv171005838T} and future campaigns, the light curves to be observed in X-rays at 1 keV, optical band at 1 eV and radio wavelength at 6 GHz were derived for the shock-breakout material and the relativistic off-axis jet. We found that fluxes have similar behaviors depending on which power-law segment of the spectrum are evolving. We have shown that they  are strongly dependent on the values of $p$ and $\alpha_s$; at early times  fluxes, in general, are dominated by those generated with small values of $\alpha_s$  whereas at later times are dominated by those with larger values.\\
In a particular case, we have considered the multiwavelength afterglow observations detected from GW170817 and found the best-fit values of a set of 8 parameters, $\Xi_{\rm fow, b}$ = \{ $\tilde{E}$, n, p, $\theta_j$, $\Delta \theta$, $\varepsilon_B$, $\varepsilon_e$, $\alpha_s$ \} for our afterglow model using the MCMC method.   For the shock-breakout material, we found that  the bulk Lorentz factor becomes $\Gamma_{\rm c}\simeq 3.1 \left(t/15\,d\right)^{-0.24}$ and  the equivalent kinetic energy $\simeq 3.31\,\times 10^{47}\,{\rm erg}$ which corresponds to an efficiency of $\sim 16\%$.   The cooling spectral break $\epsilon_{\rm c}\sim19.2$ keV is above the X-ray band and the characteristic break $\epsilon_{\rm m}=1.2\times10^{-3}$ GHz is below the radio band at 15 days. The X-ray, optical and radio fluxes peak at $\sim$ 30 days, and later they evolve as $F_\nu\propto t^{-0.76}$. At $\sim$ 500 s, the X-ray flux will vary from $\propto t^{-0.76}$ to  $\propto t^{-1.03}$ whereas  the optical and radio fluxes will continue evolving as $F_\nu\propto t^{-0.76}$.   For the relativistic jet, we found that  the bulk Lorentz factor becomes $\Gamma_{\rm j}\simeq 4.6 \left(t/140\,d\right)^{-\frac32}$.   The cooling spectral break $\epsilon_{\rm c}\sim3.5$ keV is above the X-ray band and the characteristic break $\epsilon_{\rm m}=0.03$ GHz is below the radio band at 100 days.  During this period, the observed flux increases as $F_\nu\propto t^{4.2}$ (eq. \ref{scsyn_t}), as predicted in \cite{2018arXiv180109712N}.  The X-ray, optical and radio fluxes peak at $\sim$ 140 days, and later they evolve as $F_\nu\propto t^{-2.2}$. \\  
\\
Recently, \cite{2018arXiv180609693M} reported new detections in radio wavelengths collected with Very Long Baseline Interferometry (VLBI).  These observations exhibited for almost 150 days (between  75 and 230 days post-merger) superluminal motion  with apparent speed of $\sim$ 4. This provided direct evidence that binary NS  system in GW170817 launched a relativistic narrowly collimated jet with an opening angle less $\lesssim 5^\circ$, a bulk Lorentz factor of $\sim$ 4 (at the time of measurement), observed from a viewing angle of $20^\circ\pm5^\circ$.   Our model is consistent with the results shown by  \cite{2018arXiv180609693M}, which the earlier emission is dominated by the slower shock-breakout material, and the later emission ($\gtrsim$ 80 days post-merger) by a relativistic off-axis jet.  Taking into account the values of $\Delta \theta\simeq16^\circ$ and $\theta_j\simeq 5^\circ$ reported in Table \ref{table5:param_late}, the value of the viewing angle $\theta_{\rm obs} \sim 21^\circ$ is found, which agrees with that reported in \cite{2018arXiv180609693M}.   The observed flux generated by the deceleration of the mildly-relativistic shock-breakout material dominates at early times ($\lesssim$ 50 days) and the relativistic off-axis jet dominates at later times ($\gtrsim$ 80 days). This behaviour its due to the fact that the mildly-relativistic shock-breakout material is seemed on-axis ($\theta_{\rm c}\simeq 16^\circ$) whereas the relativistic jet is off-axis with $\theta_{\rm obs}>2\theta_{\rm j}$.  It is worth mentioning that the values of opening angle, the bulk Lorentz factor and the viewing angle reported by \cite{2018arXiv180609693M} also agree with those found in our model.\\
\\
The  binary NS system ejects several materials during the merger. In addition to a relativistic jet and a shock-breakout material, a dynamical ejecta and/or neutrino-driven wind are also launched.   Since the relativistic jet makes its way out inside the dynamical ejecta, the energy deposited laterally could create a cocoon.  Depending on the duration and the energy deposited by the relativistic jet in the dynamical ejecta, the cocoon will be or not formed  \citep{2014ApJ...788L...8M, 2014ApJ...784L..28N, 2017ApJ...834...28N, 2017MNRAS.471.1652L, 2017ApJ...848L...6L}.  If the relativistic jet is launched before the ejecta begins to expand, then its propagation through the dynamical ejecta cannot inflate a cocoon and hence it will be neglected \citep{2018MNRAS.473..576G}.   We argue that in GW170817  there was no delay between the explosion (i.e., the ejection of the shock breakout ejecta) and the ejection of the relativistic jet, so the cocoon emission is neglected.    In this previous case the delay of $1.74\pm0.05$ s\citep{2041-8205-848-2-L12} found between the NS merger GW chirp signal and the $\gamma$-ray flux detected by GBM could be interpreted in terms of the  extra path length that radiation travels from the edge of the off-axis  jet to an observer in comparison with the GW which is emitted in the observer's direction.  This geometrical delay expressed in terms of $\Delta\theta =16^\circ$ and   
the distance from the central engine to the emitting region $R_\gamma$  is given by \citep[see Figure 1 shown in][]{2017ApJ...850L..24G}
\be
t_{\Delta\theta}= R_\gamma[1-\cos\Delta\theta]\simeq 1.71\,{\rm s}\,R_{12.1}\,.
\ee
It is worth noting that although the jet moves at speed slightly less than the speed of light from the acceleration phase to the internal shocks take place $R_\gamma$, this extra delay is neglected $t_\gamma\simeq R_\gamma \Gamma^{-2}_{\rm j}\simeq 4.6\times 10^{-3}\,{\rm s}\,R_{\gamma,12.1} \Gamma^{-2}_{\rm j,2}$. Therefore, the observed delay between the GW signal and the $\gamma$-ray flux can be explained in the framework of  a geometrical delay $t_{\Delta\theta}$.\\
\cite{2001ApJ...551..946T} studied a transrelativistic acceleration model, in the context of a supernova explosion, and modelled the kinetic energy of the outer material expelled. These authors found that the equivalent kinetic energy of the outermost material could be described through a power-law velocity distribution and also showed that part of it would be given to the circumburst medium, generating a strong electromagnetic emission.  \cite{2014MNRAS.437L...6K} applied the transrelativistic acceleration model to describe the shock-breakout material ejected in the binary NS merger.   They showed the kinetic energy distribution of the shock-breakout material  for different polytropic indexes $n=$3, 4 an 6 and masses ejected by the  shock breakout $M_{\rm sh}=10^{-4},\,10^{-5}$ and $10^{-6}\,M_{\rm sun}$. Given the values reported in Table \ref{table5:param_late}, the equivalent kinetic energy is $E_{\rm k, c}(\gtrsim\beta_{\rm c} \Gamma_{\rm c}) \simeq  3\times 10^{49}\,{\rm erg}  \,\left(\beta_{\rm c}\Gamma_{\rm c}\right)^{-2.3}=3.1\times 10^{48}\,{\rm erg}$, for $\Gamma_{\rm c}\simeq 3$. This value agrees with the most optimistic scenario reported in Figure 2  (n=3 and $M_{\rm sh}=10^{-4}\, M_{\rm sun}$) by  \cite{2014MNRAS.437L...6K}.   It is worth noting that although values of higher kinetic energies  are difficult to resolve, by recent 3D merger simulations, relevant implications for the NS equation of state must be analyzed with caution.\\
\\
\cite{2017ApJ...848L..20M} and \cite{2017ApJ...848L..21A} studied the X-ray and radio light curves  of GRB 170817A  in a context of  standard synchrotron emission  from the forward-shock model.  Authors  concluded that the on-axis afterglow emitted by a jet was ruled out arguing that although this model can describe  the X-ray light curve, it fails to comply the upper limits in the radio light curve which varies as $\propto t^{\frac12}$.   We propose that the X-ray, optical and radio fluxes are not emitted from a decelerated jet but from the fraction of the outermost layer moving towards us which evolves with a steeper slope $F_\nu\propto t^{0.64}$ before the break.  The evolution of this model since few days are below the upper limits and consistent with the observations.\\
\\
The dynamics of different  masses ejected from the merger with significant kinetic energies has been investigated as possible electromagnetic emitters \citep{2013ApJ...778L..16H, 2014MNRAS.437L...6K, 2015MNRAS.450.1430H}.  Authors suggested that the electromagnetic signatures associated with the deceleration of these relativistic and subrelativistic masses by the circumburst medium could be detected from $\gamma$-rays to radio wavelengths and could be observed at nearly all the viewing angles.   For instance,   \cite{2014MNRAS.437L...6K} considered a power law velocity distribution with $E\propto \Gamma^{-1.1}$ for $n_p=3$ and  proposed that the shock-breakout material ejected at ultrarelativistic velocities ($\Gamma\simeq40 - 400$) could be decelerated  emitting early photons by synchrotron radiation which would be  detected in current X-ray and radio instruments.  \cite{2015MNRAS.450.1430H} studied the dynamics and the radio components emitted by different ejected masses including a dynamical ejected mass and a cocoon.  Assuming a breakout material ejected at subrelativistic velocities $\beta_{\rm c}\Gamma\simeq 1$ for a velocity distribution of $E_{\rm k,c}\propto (\beta_{\rm c}\Gamma_{\rm c})^{-3}$ for $n_p=3$, authors showed that in all cases an early electromagnetic component from the decelerated material is expected at the radio wavelengths.   The light curves derived in this paper are different to those derived in the above papers.   In this paper,  we derive the synchrotron light curves generated from the mildly relativistic shock-breakout material and the relativistic off-axis jet when both are decelerated by an homogeneous density for the adiabatic index  $n_p=3$. We have shown that at early times before 50 days, the emission originated from the decelerated shock-breakout material dominated and at later times larger than $\gtrsim$ 80 days the emission is dominated from the off-axis jet. For instance, X-ray flux derived in \cite{2014MNRAS.437L...6K} only decrease with time as $\propto t^{-0.42}$, and in our case X-ray fluxes increase as $\propto t^{0.64}$ at early times. It is worth noting that masses ejected from a collapsar scenario to describe  low-luminosity GRBs also have been considered \cite[e.g. see;][]{2015MNRAS.448..417B}. \\ 
\\
While writing this paper we became aware of a recent preprint \citep{2018arXiv180109712N} which explains that the increase in  X-ray and radio flux  observed in GW170817 could be explained in terms of  the synchrotron radiation originated in a decelerated material moving at larger angles. From the observations they obtained values of  the bulk Lorentz factor and the isotropic equivalent energy similar to those reported in this paper for GW170817 event.\\
\\
Electromagnetic counterpart observations  from a binary NS system associated with gravitational waves cast the merger scenario in new light.  Similar analysis to the one developed in this paper with futures observations can shed light on the properties of the outer ejected materials. \\
\section*{Acknowledgements}
We thank Enrico Ramirez-Ruiz, Jonathan Granot, Dafne Guetta, Fabio de Colle, Rodolfo Barniol-Duran and Diego Lopez-Camara for useful discussions.  NF  acknowledges  financial  support from UNAM-DGAPA-PAPIIT through grants IA102917 and IA102019. ACCDESP acknowledges that this study was financed in part by the Coordenação de Aperfeiçoamento de Pessoal de N\'ivel Superior - Brasil (CAPES) - Finance Code 001 and also thanks the Professor Dr. C. G. Bernal for tutoring and useful discussions.  PV  thanks  Fermi  grants NNM11AA01A and 80NSSC17K0750.   
%
%
%

%
\clearpage
\begin{figure}
{ \centering
\resizebox*{1.\textwidth}{0.9\textheight}
{\includegraphics{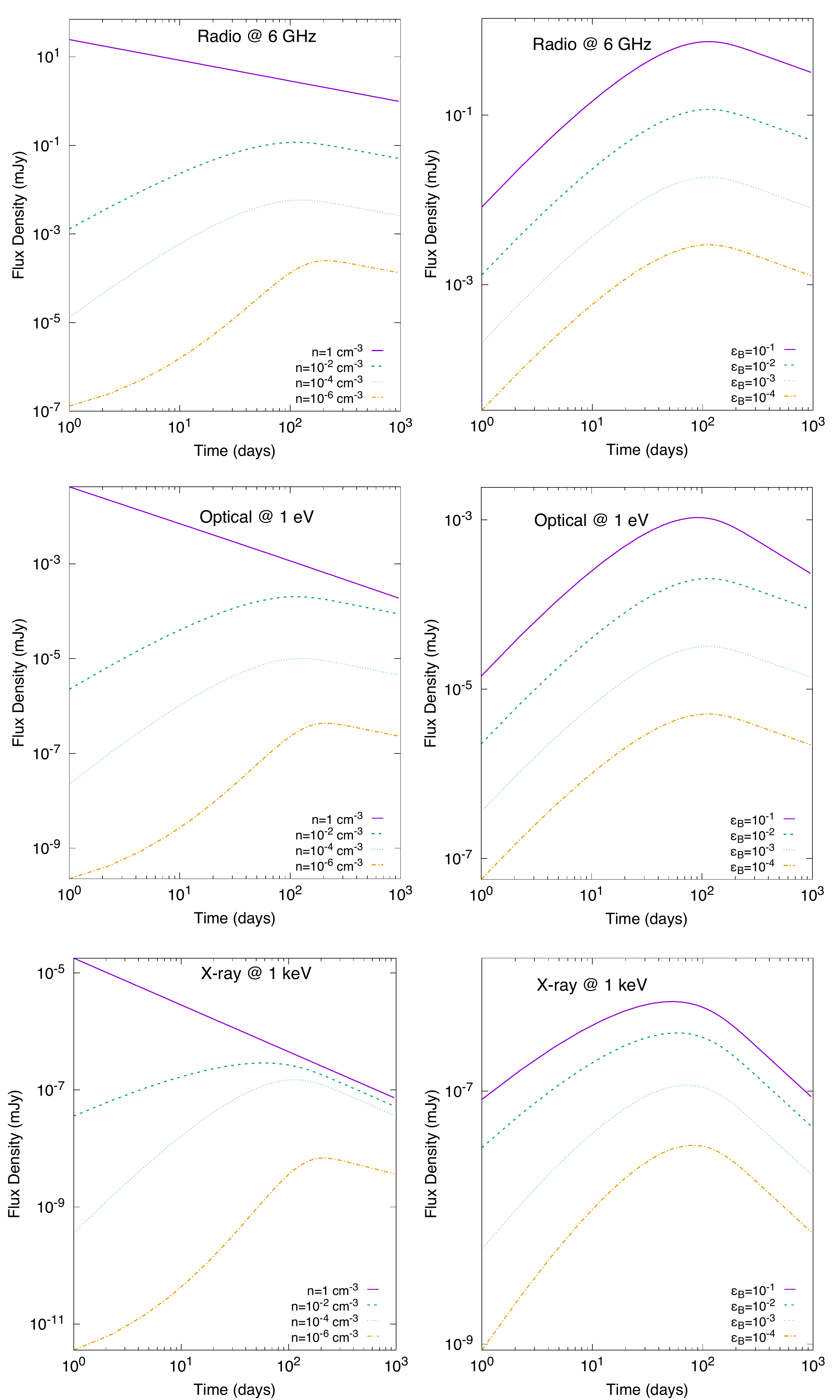}}
}
\caption{Radio (upper), optical (medium) and X-ray (lower) light curves of the synchrotron radiation emitted from the deceleration of  the shock-breakout material. The left- and right-hand panels exhibit the light curves considering the magnetic microphysical parameter $\epsilon_B=10^{-2}$ and  density $n=10^{-2}\,\,{\rm cm^{-3}}$, respectively,  for the values of fiducial energy $\tilde{E}=10^{50}$ erg, luminosity distance $D$=100 Mpc and indices $p=2.2$,  $\alpha_s=3.0$ }
\label{lc1}
\end{figure}

\clearpage
\begin{figure}
{ \centering
\resizebox*{1.\textwidth}{0.85\textheight}
{\includegraphics{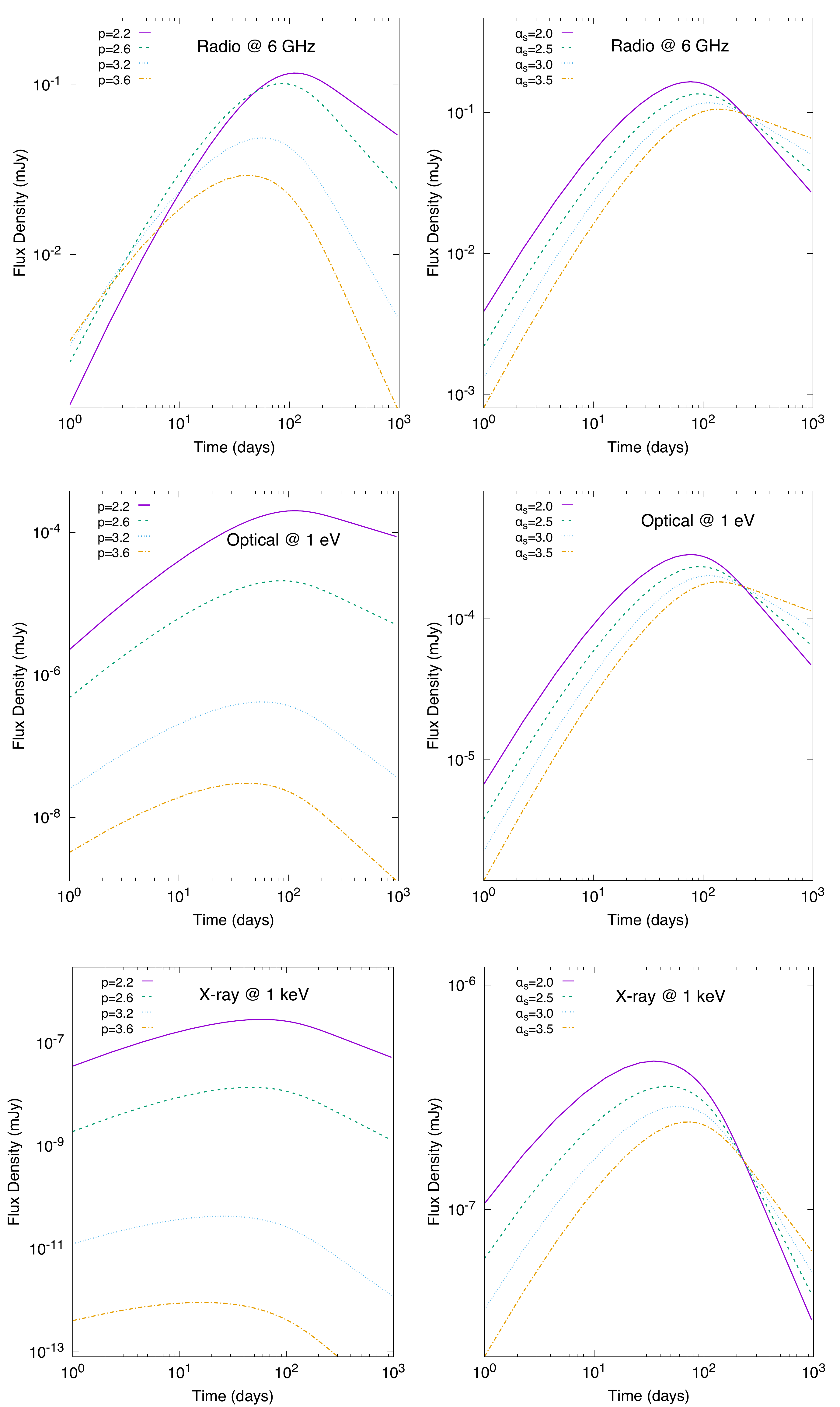}}
}
\caption{Radio (upper panel), optical (medium panel) and X-ray (lower panel) light curves  of the synchrotron radiation emitted from the deceleration of  the shock-breakout material.  The left- and right-hand panels exhibit the light curves considering the indices  $\alpha_s=3.0$  and  $p=2.2$, respectively,  for the values of fiducial energy $\tilde{E}=10^{50}$ erg,  luminosity distance $D$=100 Mpc, $n=10^{-2}\,\,{\rm cm^{-3}}$,  $\epsilon_B=10^{-2}$ }
\label{lc2}
\end{figure}
\clearpage
\begin{figure}
{ \centering
\resizebox*{1.\textwidth}{0.9\textheight}
{\includegraphics{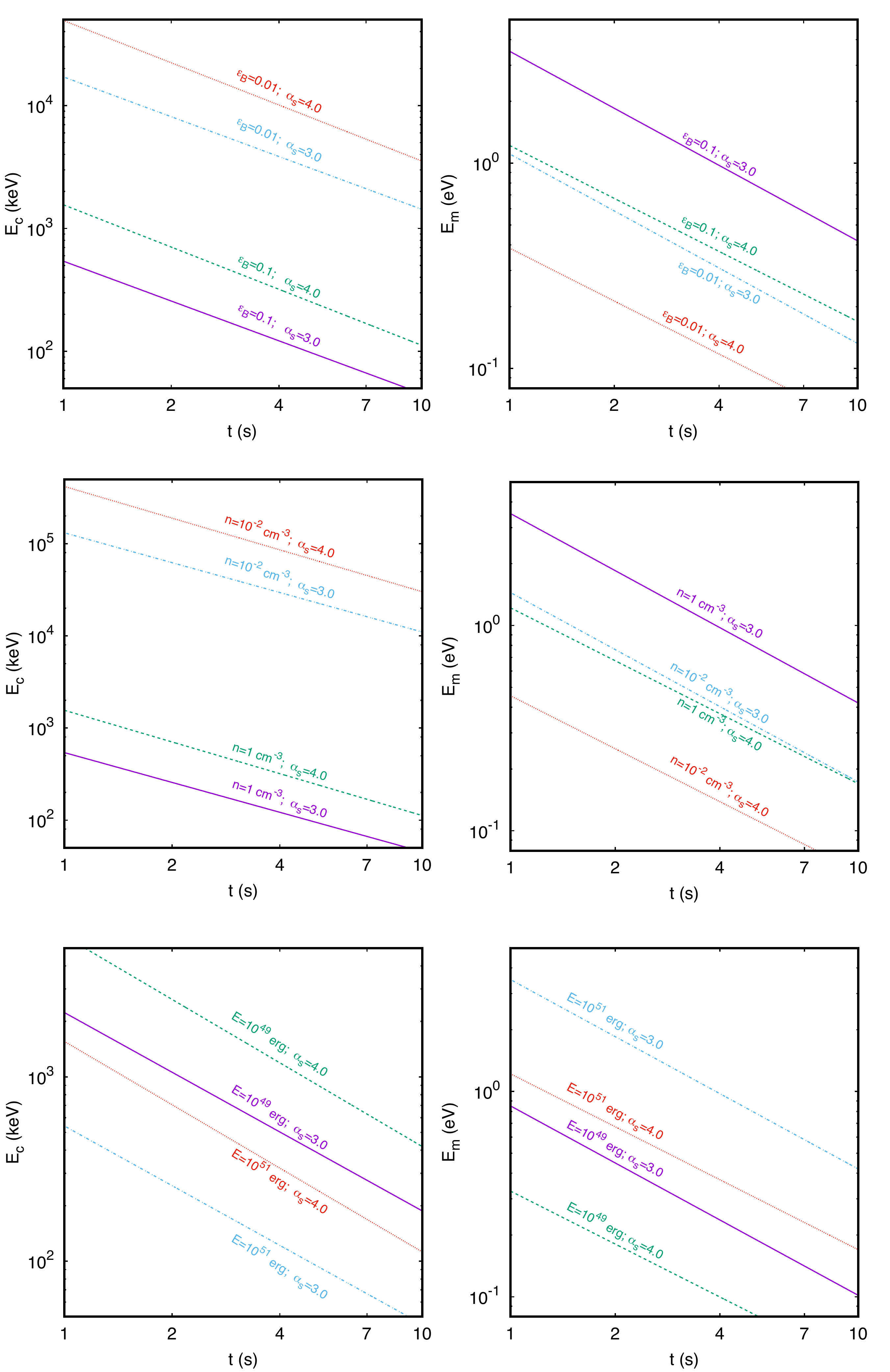}}
}
\caption{Evolution of synchrotron cooling (left-hand panels) and characteristic (right-hand panels) spectral breaks  as a function of time for typical values in ranges   $10^{-2}<\epsilon_B<10^{-1}$ (upper panel),  $10^{-2}<n<1\,{\rm cm^{-3}}$ (medium panel) and   $10^{49} <\tilde{E}<10^{51}\, {\rm erg}$ (lower panel) considering $3.0<\alpha_s<4.0$ and $p=2.2$.}
\label{Epeak}
\end{figure}
\clearpage
\begin{figure}
	{ \centering
		\resizebox*{\textwidth}{0.7\textheight}
		{\includegraphics[angle=-90]{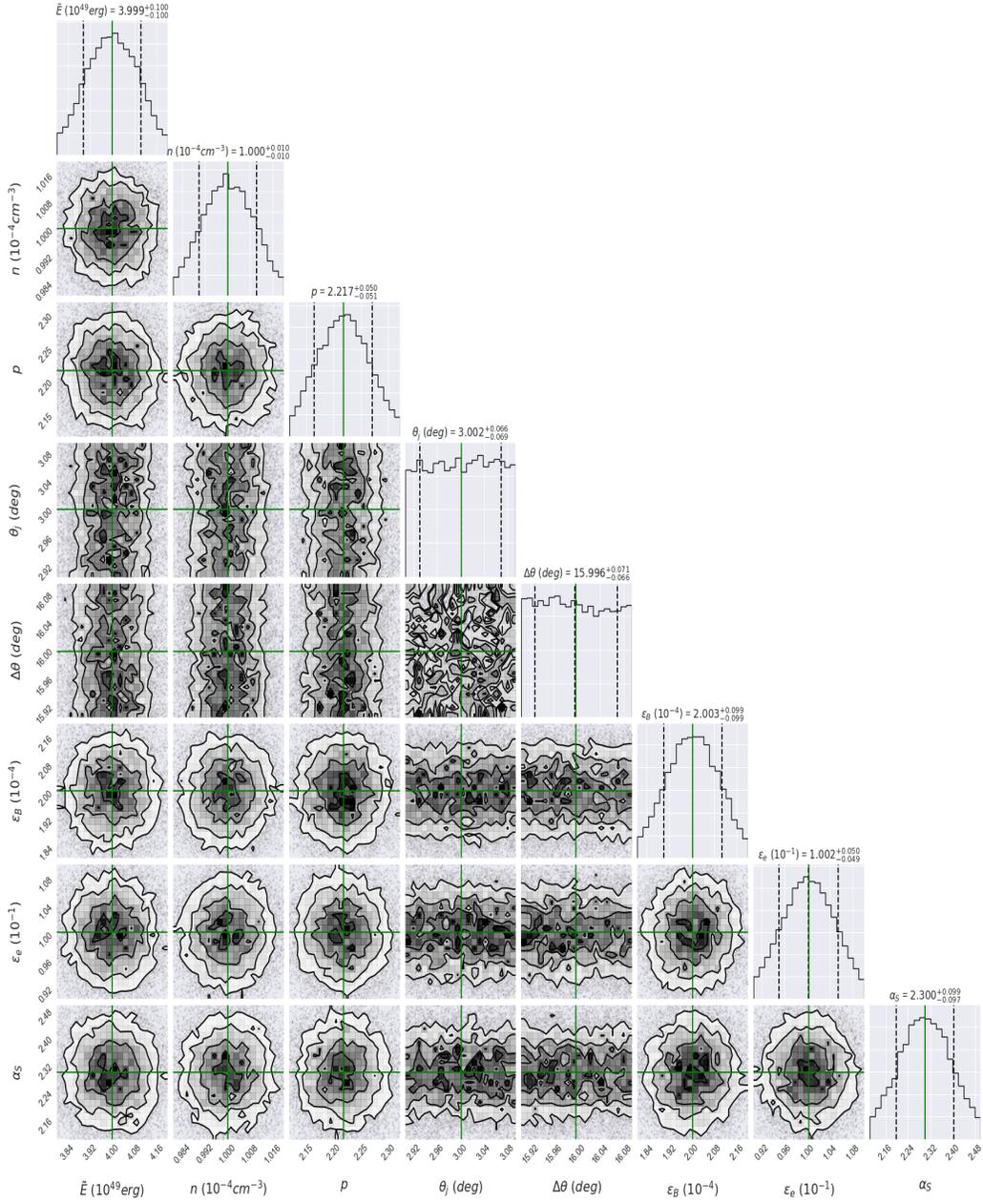}}
	}
	\caption{Corner plot demonstrating the results obtained from the MCMC simulation for our parameter set.  Fit results for the radio light curve at 3 GHz using the synchrotron forward-shock model described in Section 2 and 3 .  Labels above the 1-D KDE plot indicate the quantiles chosen for each parameter.   The best-fit value is shown in green.  Values are reported in Table \ref{table5:param_late} (Col 2).}
	\label{fig3:param_late_3R}
\end{figure}

\clearpage

\begin{figure}
	{ \centering
		\resizebox*{\textwidth}{0.7\textheight}
		{\includegraphics[angle=-90]{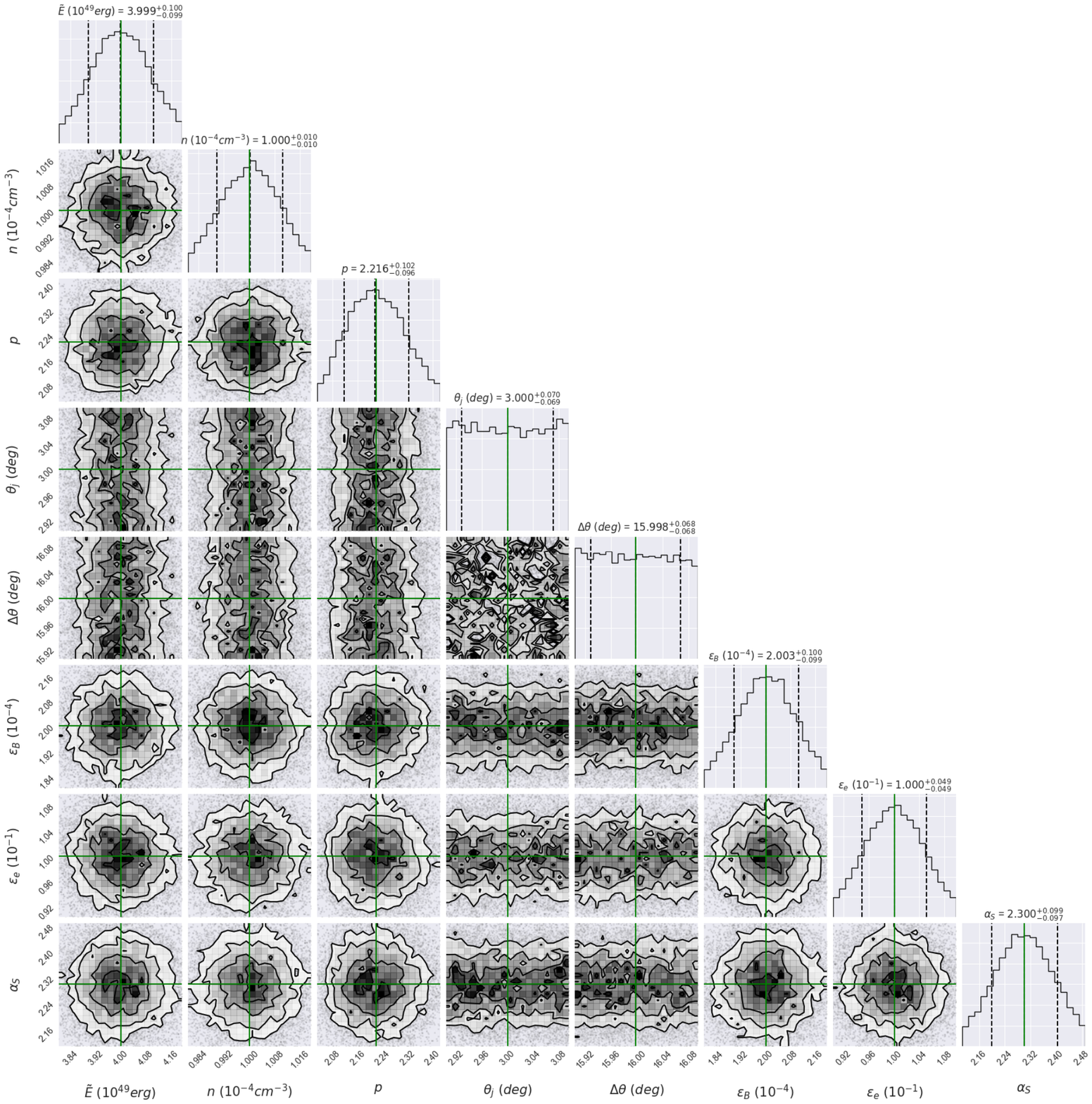}}
	}
	\caption{Same as Fig. \ref{fig3:param_late_3R}, but it shows the fit results for the radio light curve at 6 GHz. Values are reported in Table \ref{table5:param_late} (Col 3).}
	\label{fig3:param_late_6R}
\end{figure}

\clearpage

\begin{figure}
	{ \centering
		\resizebox*{\textwidth}{0.7\textheight}
		{\includegraphics[angle=-90]{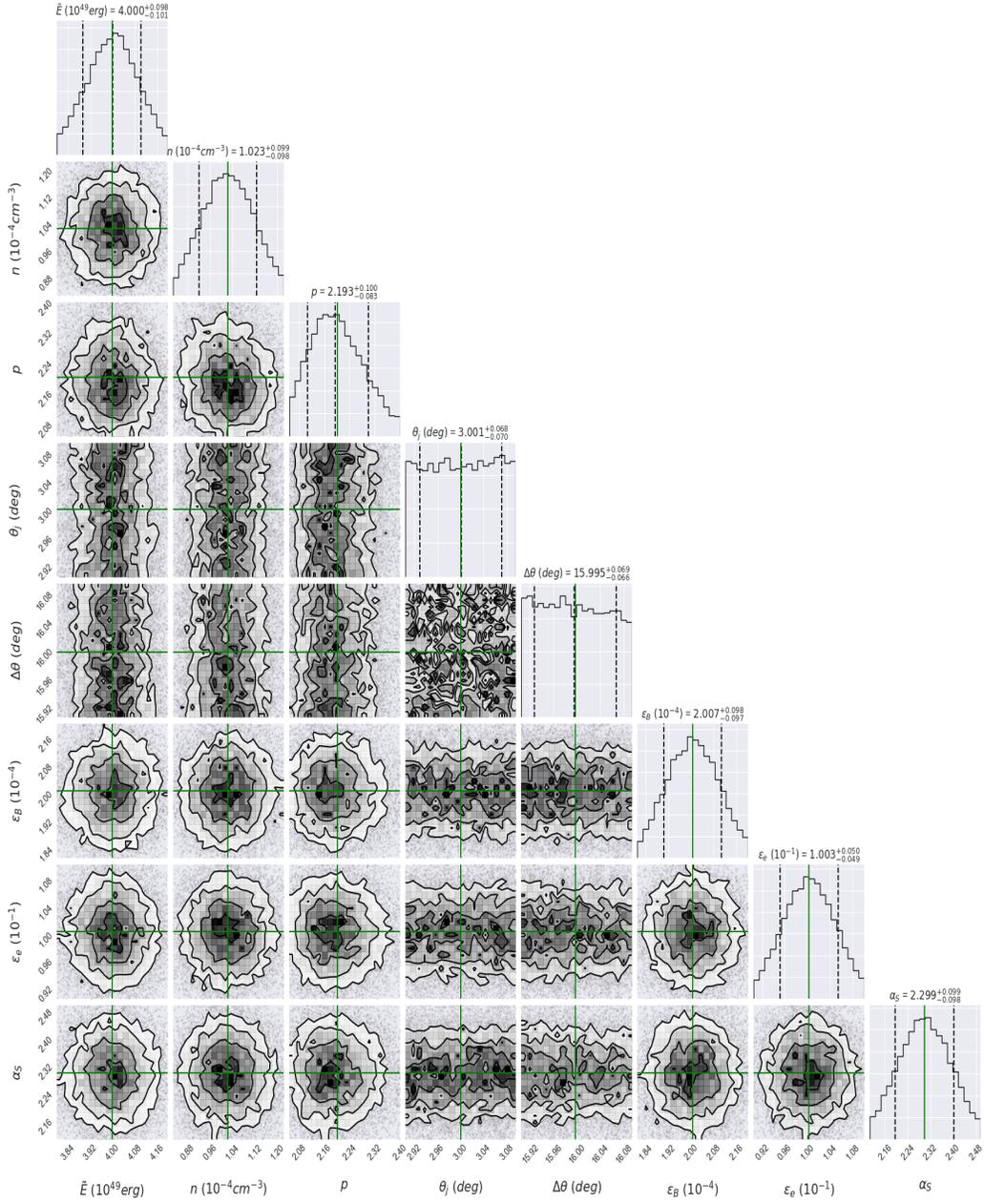}}
	}
	\caption{Same as Fig. \ref{fig3:param_late_3R},  but it shows the fit results for the X-ray light curve.  Values are reported in Table \ref{table5:param_late} (Col 4).}
	\label{fig3:param_late_X}
\end{figure}

\clearpage

\begin{figure}
{ \centering
\resizebox*{0.55\textwidth}{0.4\textheight}
{\includegraphics{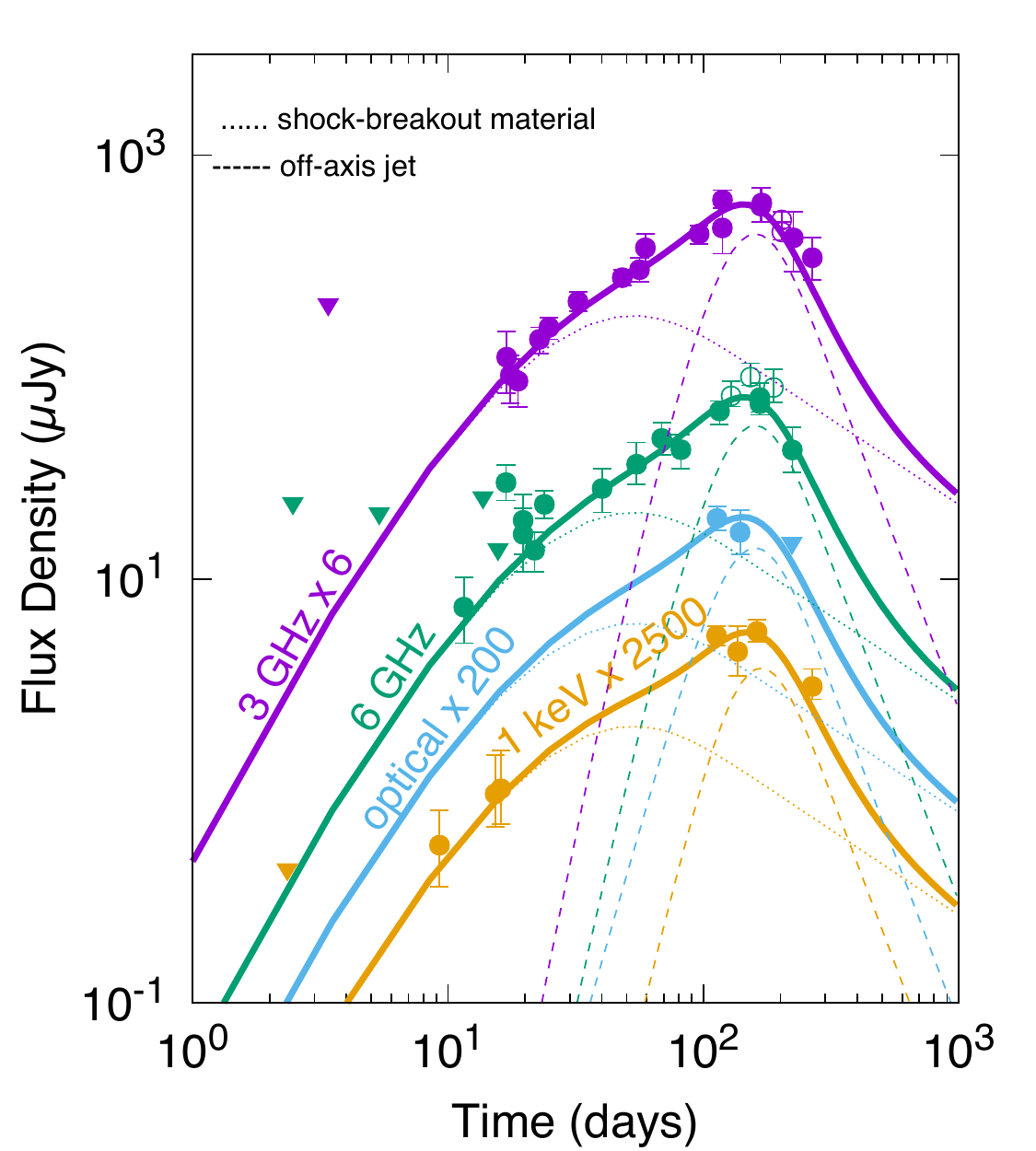}}
\resizebox*{0.5\textwidth}{0.4\textheight}
{\includegraphics{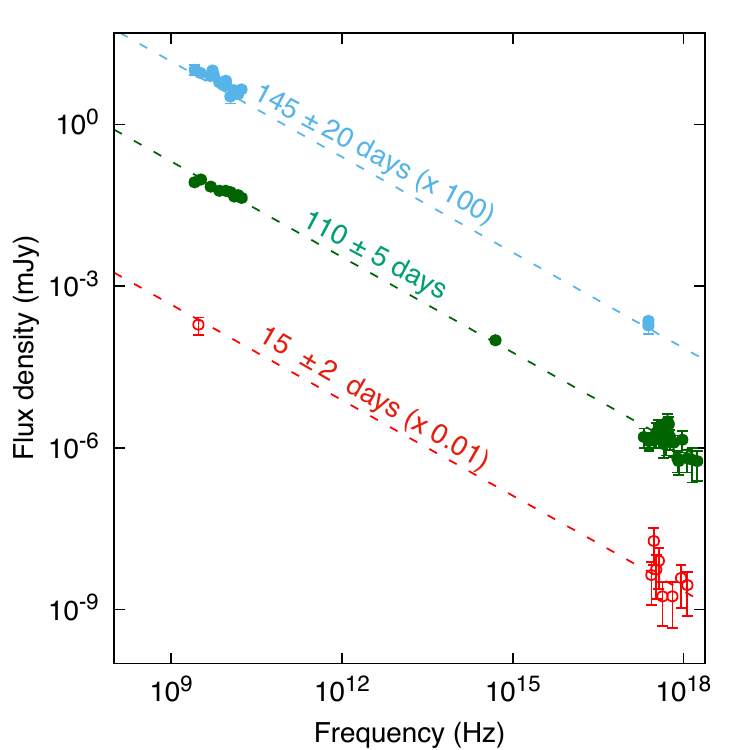}}
}
\caption{Left: Light curves of X-ray at 1 keV  \citep[gold;][]{2017Natur.547..425T, 2017ApJ...848L..20M, 2017ATel11037....1M, 2018ATel11242....1H,  troja2017a, 2018arXiv180106516T,2018arXiv180103531M},  optical  \citep[blue;][]{2018arXiv180103531M},  and radio  at 3 and 6 GHz \citep[magenta and green;][]{2017Natur.547..425T, 2017arXiv171005435H, 2017arXiv171111573M, 2017ApJ...848L..21A}   bands.  Right: SEDs of the X-ray, optical and radio afterglow observations at 15 $\pm$ 2 (red),  110 $\pm$ 5 (green) and 145 $\pm$ 20 (blue) days. The values found after describing the light curves and SED are reported in Table \ref{table5:param_late}.}
\label{fig4:afterglow}
\end{figure}
\clearpage
\begin{figure}
{ \centering
\resizebox*{0.6\textwidth}{0.4\textheight}
{\includegraphics{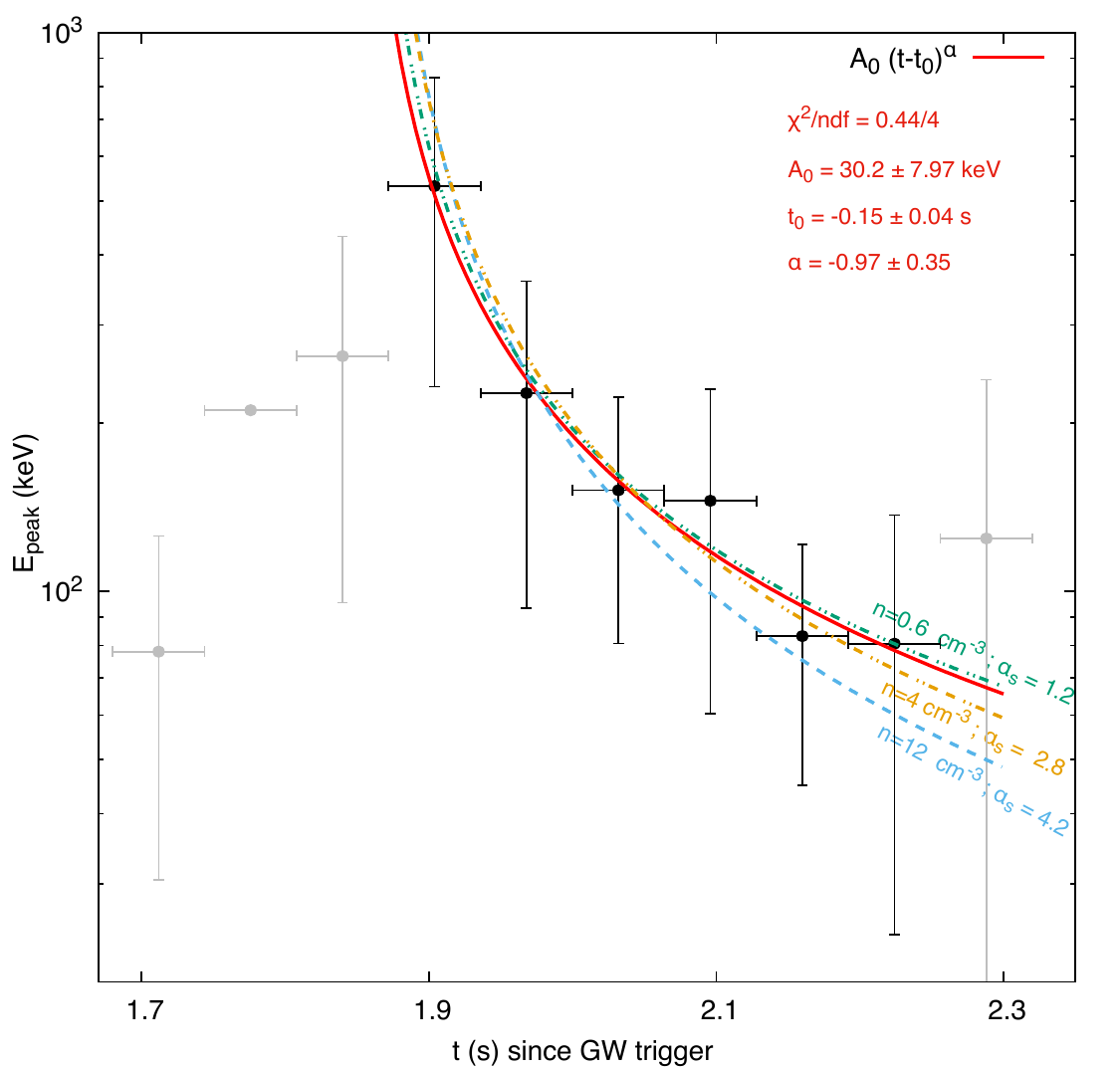}}
}
\caption{Peak energy as a function of time from the GW trigger. The red solid line is the fitted simple power law $F=A_0 (t-t_0)^{\alpha}$ with  $A_0=30.2\pm 7.97\, {\rm keV}$, $t_0=-0.15\pm0.04\, {\rm s}$, $\alpha=-0.97\pm0.35$, $\chi^2/{\rm ndf}=0.44/4$. Dashed, dotted and dashed-dotted lines represent the cooling spectral break of our theoretical model for $\alpha_s=1.6$ and $n= 1\,{\rm cm^{-3}}$ (green line), $\alpha_s=2.8$ and $n= 4\,{\rm cm^{-3}}$ (gold line) and $\alpha_s=4.2$ and $n= 12\,{\rm cm^{-3}}$ (blue line), respectively. The values used are  $\tilde{E}=3\times10^{49}\,\,{\rm erg}$, $\epsilon_B=10^{-1}$, $\epsilon_e=10^{-1}$ and $p=2.2$}
\label{Epeak_data}
\end{figure}
\clearpage
\clearpage
\begin{table}
\centering \renewcommand{\arraystretch}{2}\addtolength{\tabcolsep}{3pt}
\caption{Median and symmetrical 35\% quantiles (0.15, 0.5, 0.85), truncated at the third decimal,   after describing the X-rays and radio wavelengths at 3 and 6 GHz with our afterglow model  as described in Sections 2 and 3.}\label{table5:param_late}
\begin{tabular}{ l  l  l  l c }
\hline
\hline

{\large   Parameters}	& 		& {\large  Median}  & 		 				                           		 	        	   		 \\ 
                          	& 	{\normalsize Radio (3 GHz)} 	&  {\normalsize Radio (6 GHz)} & 		{\normalsize X-ray (1 keV)}			                           		 	        	   		 \\ 

\hline \hline
\\
\small{$\tilde{E}\, (10^{49}\,{\rm erg})$}	\hspace{1cm}&   \small{$3.999^{+0.100}_{-0.100}$}	 \hspace{0.7cm}	&  \small{$3.999^{+0.100}_{-0.099}$}	\hspace{0.7cm} &  \small{$4.000^{+0.098}_{-0.101}$}	 \\
\small{${\rm n}\,\, (10^{-4}\,{\rm cm^{-3}}$ ) }\hspace{1cm}	&  \small{$1.000^{+0.010}_{-0.010}$} 	 \hspace{0.7cm}&  \small{$1.000^{+0.010}_{-0.010}$} 	\hspace{0.7cm}&  \small{$1.023^{+0.099}_{-0.098}$}	         \\
\small{${\rm p}$}	\hspace{1cm}&  \small{$2.217^{+0.050}_{-0.051}$}  \hspace{0.7cm}&  \small{$2.216^{+0.102}_{-0.096}$} \hspace{0.7cm}&  \small{$2.193^{+0.100}_{-0.083}$}	 	                                \\
\small{$\theta_j$\,({\rm deg})}	\hspace{1cm}&  \small{$3.002^{+0.066}_{-0.069}$}	 \hspace{0.7cm}&  \small{$3.000^{+0.070}_{-0.069}$}	\hspace{0.7cm} &  \small{$3.001^{+0.068}_{-0.070}$}	 	                 \\
\small{$\Delta \theta$\,({\rm deg})}	\hspace{1cm}&  \small{$15.996^{+0.071}_{-0.066}$}	 \hspace{0.7cm}&  \small{$15.988^{+0.068}_{-0.068}$}	\hspace{0.7cm} &  \small{$15.995^{+0.069}_{-0.066}$}	 	                 \\
\small{$\varepsilon_B\,\,(10^{-4})$}   \hspace{1cm}&\small{$2.003^{+0.099}_{-0.099}$}  \hspace{0.7cm}&\small{$2.003^{+0.100}_{-0.099}$} \hspace{0.7cm}&\small{$2.007^{+0.098}_{-0.097}$}	 	                           \\
\small{$\varepsilon_{e}\,\,(10^{-1})$}	 \hspace{1cm}&\small{$1.002^{+0.050}_{-0.049}$}   \hspace{0.7cm}&\small{$1.000^{+0.049}_{-0.049}$} \hspace{0.7cm} &\small{$1.003^{+0.050}_{-0.049}$}	 	    \\
\small{${\alpha_s}$}  \hspace{1cm}&\small{$2.300^{+0.099}_{-0.097}$}   \hspace{0.7cm}&\small{$2.300^{+0.099}_{-0.097}$} \hspace{0.7cm} &\small{$2.299^{+0.099}_{-0.098}$}	 	    \\

\hline
\end{tabular}
\end{table}

\end{document}